\documentclass[showpacs,amsmath,amssymb,preprint]{revtex4}
\usepackage{graphicx}
\usepackage{subfigure}
\usepackage{hyperref}
\usepackage{listings}
\usepackage{exscale}
\usepackage{array}
\usepackage{color}
\begin{document}
\lstset{language=C++}
\title{Highly optimized simulations on single- and multi-GPU systems of 3D Ising spin glass}
\author{M. Lulli$^1$}
\email{matteo.lulli@gmail.com}
\author{M. Bernaschi$^2$}
\author{G. Parisi$^{1,3}$}

\address{$^1$ Dipartimento di Fisica di ``Sapienza", Universit\`a di Roma, 00185 Roma, Italy}
\address{$^2$ Istituto per le Applicazioni del Calcolo, IAC-CNR, Rome, Italy}
\address{$^3$ INFN, Sezione di Roma I, 00185 Roma, Italy}

\begin{abstract}
  We present a highly optimized implementation of a Monte Carlo (MC) simulator for the three-dimensional Ising spin-glass model with bimodal disorder, \emph{i.e.}, the 3D Edwards-Anderson model running on CUDA enabled GPUs. Multi-GPU systems exchange data by means of the Message Passing Interface (MPI). The chosen MC dynamics is the classic Metropolis one, which is purely dissipative, since the aim was the study of the critical off-equilibrium relaxation of the system. We focused on the following issues: $i)$ the implementation of efficient access patterns for nearest neighbours in a cubic stencil and for lagged-Fibonacci-like pseudo-Random Numbers Generators (PRNGs); $ii)$ a novel implementation of the asynchronous multispin-coding Metropolis MC step allowing to store one spin per bit and $iii)$ a multi-GPU version based on a combination of MPI and CUDA streams. We highlight how cubic stencils and PRNGs are two subjects of very general interest because of their widespread use in many simulation codes. Our code best performances $\sim 3$ and $\sim 5$ psFlip on a GTX Titan with our implementations of the MINSTD and MT19937 respectively.

\end{abstract}
\maketitle
\section{Introduction}
The three-dimensional Ising spin glass is a statistical mechanics model defined on a cubic lattice by the Hamiltonian
\begin{equation}
H = - \sum_{\langle ik \rangle} J_{ik}\, \sigma_i \sigma_k,
\end{equation}
where the $\sigma_i \in \{-1,+1\}$ are the so-called spin variables, the $J_{ik} \in \{-1,+1\}$ are the coupling constants (representing \emph{quenched} disorder) which are randomly drawn according to a given probability distribution $P(J_{ik})$ and the sum $\sum_{\langle ik \rangle}$ is restricted to nearest neighbouring spins. Since we deal with bimodal disorder the probability distribution reads
\begin{equation}
P(J_{ik}) = \frac{1}{2} \left [ \delta_{\scriptscriptstyle{K}}(J_{ik} - 1) + \delta_{\scriptscriptstyle{K}}(J_{ik} + 1) \right ],
\end{equation}
where $\delta_{\scriptscriptstyle{K}}(a-b)=\delta_{ab}$ stands for the Kroneker delta. Such a model describes, in three dimensions, a disordered and frustrated magnetic system showing a glassy dynamics below a finite critical temperature $T_c = 1.1019(29)$ \cite{Janus2013Eq}. Because of a very high dynamic critical exponent $z=6.86(16)$ \cite{JanusZ} for the MC Metropolis dynamics, this model has represented a long standing challenge for numerical simulations. For almost thirty years special purpose machines have been employed \cite{OgielskiSuperComputer, Cruz2001165SUE, JanusTempi, Janus2_2014} in order to get equilibrium measures for always larger systems. Large systems are needed because of the severe finite-size corrections to scaling \cite{Pelissetto2008, Janus2013Eq}. Moreover, new kinds of dynamics have been developed in order to reach equilibrium as fast as possible: Parallel Tempering has proved to be the best choice so far \cite{MarinariParisi_PT}.

However, the hegemony of special purpose hardware for this class of problems might be about to end because of the steadily increasing use of (multi) GPU devices enabling us to reach computational horsepower exceeding by orders of magnitude those of common CPUs.

Since the very early use of CUDA, and even before \cite{Tomov200571_firstGPU}, it has been understood that MC simulations of Ising spin systems would have enjoyed benefits from the use of GPUs. This is not surprising: for even cubic lattice sizes $L=2n$ the system can be simply partitioned according to a checkerboard scheme into two coloured subsystems, which we will refer to as \emph{reds} and \emph{blues}, which can be updated separately since nearest neighbours of one colour belong to the other, \emph{i.e.}, a red spin has blue nearest neighbours only.
Hence, the problem exposes an intrinsic parallelism which perfectly suits the GPU architecture: the update of each spin of a given color does not require any coordination with the update process of other spins of the same color so that the large amount of computing threads needed for the best use of the GPU can be programmed to update concurrently independent spins of the system.

Using the Metropolis dynamics, the update of the entire system is performed via two separate kernels, one for each colour. This is an easy way to enforce the independence of the update of the two subsets, a necessary condition for a correct implementation of the Markov chain. Of course, this kind of update does not ensure \emph{detailed balance} but \emph{balance} holds nonetheless, which is enough to properly carry on a MC simulation of such systems where the system probability distribution converges to the equilibrium Gibbs distribution \cite{balanceSufficient, balance_seqIsing}
\begin{equation}
  P(\{\sigma_i\}, \beta) = \frac{e^{-\beta H[\{\sigma_i\}]}}{\sum_{\{\sigma_i\}} e^{-\beta H[\{\sigma_i\}]}}.
\end{equation}

Now, we will review the previous works on spin systems for GPUs, taking into account also different models and dimensionalities, analyzing them according to:
\begin{itemize}
\item memory allocation strategies and spins-threads mappings;
\item the kind and implementation of PRNGs;
\item multi-GPU implementation techniques.
\end{itemize}
We can classify the preceding works on spin systems according to the allocation and the memory access strategies starting from the lowest level (Global Memory) up to the highest level (Shared Memory and registers) of the memory hierarchy:

\begin{enumerate}
\item \textbf{Global Memory allocation}. Two different strategies are mainly used for the memory allocation of spins in the GPU Global Memory: a first one keeping a mixed scheme where one array is allocated containing both colours in a cubic lattice topology \cite{Preis20094468,Block20101549, Weigel20111833, LevyCohen_IsraelGPU, WeigelHeisenberg2012, Fang2014}; a second one allocating two separate buffers for the two colours breaking the cubic lattice topology \cite{Tomov200571_firstGPU, Ferrero20121578, Bernaschi20111265, Bernaschi20121416, Bernaschi2013250}.

Though the first strategy seems to be more natural one has to take into account that memory transactions are served from the L2 cache in blocks of 128 bytes so that for each transaction one loads both the to-be-changed spins and the nearest neighbours which stay unchanged during the kernel execution. Different threads will update different spins sharing some neighbours thus rendering the access pattern highly non-trivial. As a matter of fact, many of the works adopting the first strategy use the Shared Memory to improve the locality of such mixed access to the coloured spins \cite{Block20101549, Weigel20111833, Fang2014, WeigelHeisenberg2012}.

In the second case there are several benefits but also drawbacks. On one hand it is possible to achieve good memory loading performances since many threads would look for the same neighbouring spin allowing for a higher second hit probability in L1/L2 caches. However, one has to deal with an algebraically demanding access pattern due to the loss of the cubic topology: it is necessary to take into account the \emph{parity} of the lattice site in order to correctly determine the right and left neighbours. Usually this strategy does not require the use of Shared Memory as reported in \cite{Ferrero20121578, Bernaschi20111265, Bernaschi20121416, Bernaschi2013250}.

Other considerations are in order. Being the two colours allocated in two different arrays it is possible to bind each buffer to a texture in order to load the neighbouring spins through the dedicated texture unit of each Streaming Multi-processor (SM) with a separated cache different from the L1/Shared. This choice offers a two-fold advantage: on one side the slowing down due to occasional non-coalescence of loads for nearest neighbours is softened because texture fetches work on a memory locality principle, on the other side the dedicated texture hardware is in charge of the computation of physical memory addresses rather than CUDA cores. 

\item \textbf{Spins arrangement in Global Memory}. For both allocation strategies it is still possible to choose several spins arrangements. Such a choice aims at maximizing the loading efficiency from the Global Memory, \emph{i.e.} reducing the number of non-coalesced loads.

In case of a single buffer for both colours two different strategies have been proposed in \cite{WeigelHeisenberg2012} and \cite{Fang2014}. In \cite{WeigelHeisenberg2012} the authors divide the cubic lattice in sublattices linearly organized in memory in a ``snake'' fashion so that every block taking charge of one sublattice could load more efficiently the spins to the Shared Memory. As for \cite{Fang2014} the authors studied a so-called \emph{shuffled} scheme where spins coming from different replicas where mixed saving memory transactions. However they found that such a strategy performs worse than the so-called \emph{unified} one where each array contains spins belonging to one replica.

In the case of a separate allocation of colours, the authors of \cite{Ferrero20121578} proposed, for the two-dimensional Potts model, a coordinate transformation of the lattice that leads to have three of the four nearest neighbours lying sequentially in the array. 

\item \textbf{Spins-threads correspondence}. The mapping between spins and threads
can also be done in different ways.
%

Some of the works adopting the unified allocation scheme resort to (per-block) Shared Memory \cite{Block20101549, Weigel20111833, Fang2014}. In one of the most commonly used mappings each thread-block evaluates the MC move on a sublattice \cite{Block20101549, Weigel20111833, LevyCohen_IsraelGPU, WeigelHeisenberg2012, Fang2014} which usually is a square in two dimensions or a cubic sublattice in three. Clearly, with this strategy one needs to look for neighbours in the boundaries which will also be retrieved by neighbouring blocks thus leading to a duplication of data served by memory transactions. The most frequently used technique, in this case, is loading in Shared Memory the neighbours and, if needed, the couplings. However, this might represent a serious limitation since the number of thread blocks running on a single Streaming Multiprocessor (SM) depends on the amount of Shared Memory needed by each of them. In the case of models with complex degrees of freedom only few blocks can be run concurrently thus leading to underuse the SM.

A rather different, but apparently less effective, approach has been tried in \cite{Preis20094468} where thread-blocks were associated to stripes of the cubic lattice of size $L\times 2\times 2$. A thread-block is associated to each region and each thread updates 4 spins in the three-dimensional case.

As for the separated scheme there are no particular restrictions on the dimensionality of the thread blocks which can also be taken as one-dimensional. Hence, no specific correspondence between blocks and lattice portions has to be considered resulting in a more tunable and flexible scheme \cite{Bernaschi20111265, Bernaschi20121416, Bernaschi2013250}. Indeed, such a choice allows to decrease memory transfers redundancy. 
Moreover, as shown in \cite{Bernaschi20111265} and verified in the present work, it turns out that nearest neighbours values are loaded in a more efficient way directly from global memory making use of texture fetches, \emph{i.e.} texture cache and hardware.

\end{enumerate}

We continue now with the choice of the PRNG which is one of the most important aspects of a MC simulation. The reliability of the estimates depends on the quality of the sequence generated by the chosen PRNG. As an example, it is well known that the use of one single Linear Congruential PRNG with a period $p = 2^k$ in equilibrium MC for the 2D Ising model leads to systematic discrepancies on lattice sizes $L=2^\ell$ due to resonance phenomena between the size of the system and the systematic long-range correlations which affect Congruential PRNGs. It is then important to provide fast implementations for reliable PRNGs. Nonetheless, many of the previous works on MC simulations of spin systems on GPU used such PRNGs \cite{Tomov200571_firstGPU, Preis20094468, Weigel20111833, LevyCohen_IsraelGPU, Bernaschi20111265, Bernaschi20121416, Bernaschi2013250}, mainly for benchmarking reasons. The main reason is that the state of these PRNGs is limited to one integer value making them the best choice in terms of speed but a questionable choice as for the quality of the produced numbers. However, it is interesting to notice that in \cite{Weigel20111833} compatible results with theoretical predictions for the two-dimensional Ising model were reported. In the most straightforward implementation of Linear Congruential PRNGs, each thread accesses its own memory location storing the one-valued state of the generator so that one deals with a battery of generators rather than a single one used to update the entire lattice. Little is known about the behaviour of such parallelized implementations of Linear Congruential PRNGs and it would be interesting to study them carefully.

In \cite{Ferrero20121578, WeigelHeisenberg2012} the authors used the so-called multiply-with-carry PRNG which consists in a modified Linear Congruential PRNG where the result of the module operation is used for the successive update hence requiring two integers to store the state.

In \cite{Fang2014} the cuRand implementation of the XORWOW has been used. This PRNG consists in a XOR-Shift summed to a Weyl generator.

There are a few other works using the so-called lagged-Fibonacci PRNG \cite{a0_25_PhysRevB.88.144104, MarcoBaity2013}: these generators use a state of a certain length from which two `lagged' entries are read and combined, usually summed, giving the random number and updating at the same time the state. Usually, this kind of generators have very long periods, much longer with respect to Linear Congruential PRNGs. As we are mainly interested in the memory access scheme for the GPU implementation we can label as lagged-Fibonacci-like all those PRNGs sharing a scattered read pattern of the state. Since there are lags between the reads of the state, a certain amount of random numbers can be produced in parallel by different threads \cite{WarpGeneratorSite, WeigelPRNG2012} using Shared Memory to store the state which will be used in a thread block. One of the most popular generators of this kind is the Mersenne Twister MT19937 \cite{MatsumotoMT} which has a very long period $p=2^{19937} - 1$. However, since its state needs at least 624 entries, a Shared Memory implementation would be too memory-consuming thus strongly limiting the SM occupancy. Hence in \cite{a0_24-1742-5468-2013-09-P09001} the authors chose to use the so called Warp generator \cite{WarpGeneratorSite} which has been written along the same lines of MT19937.

To our knowledge there is only one work \cite{MarcoBaity2013} using the so-called Parisi-Rapuano generator \cite{ParisiRapuano} which basically consists in a lagged-Fibonacci PRNG. However no implementation details are given. Indeed, in \cite{WeigelPRNG2012} it has been observed that such a PRNG is not well suited for a Shared Memory-based GPU implementation because of the lags values.

Finally, as for multi-GPU implementations exposing strong scaling, we are only aware of \cite{Block20101549, Bernaschi2013250, Bernaschi20121416}. Other works present weak scaling \cite{MarcoBaity2013, Fang2014} although in \cite{MarcoBaity2013} communication between different nodes is needed because of the adopted Parallel Tempering implementation. However, the requirement of strong scaling depends on the physical features of the simulated systems one is interested in.

The outlook of the paper is the following. In section 2 we describe a new cubic-stencil access pattern which uses a separated allocation scheme while keeping a cubic lattice topology and avoiding the use of Shared Memory. This result is obtained via a suitable spin arrangement which follows from a coordinate transformation akin to \cite{Ferrero20121578}. Texture fetches are used to load both nearest neighbours and couplings. Different spins-threads mappings have been tested and results are compared to those obtained with an access pattern proposed in \cite{Bernaschi20111265} handling only $L=2^{\ell}$ sizes and its generalization to $L=2n$. In section 3 we present a new implementation of the Parisi-Rapuano PRNG which avoids the Shared Memory. In order to show the advantages of this approach we will propose a GPU implementation of a per-thread version of the MT19937, the first one in our knowledge, comparing its performances with the device API cuRand MTPG32 (a GPU-tailored variant of the Mersenne Twister) and a very recent version of the host API cuRand MT19937. At the same time we propose a simple benchmarking test for the device API whereas for the host API we comply with the benchmark proposed in \cite{PRAND}. In section 4 we describe a new asynchronous multispin-coding technique and its implementation. We then show the results obtained on a variety of different GPUs and analyze them. In section 5 we present the strongly scaling multi-GPU version of the code outlining its features showing the results obtained on the Piz Daint \cite{daint} supercomputer. In section 6 we draw conclusions.

\section{Cubic Stencil}
With ``Cubic Stencil'' we refer to a set comprising one vertex of a cubic lattice together with its six nearest neighbouring vertices and edges. It represents the set of data needed to perform the update of a spin. It is also the fundamental data element of many other algorithms based on the cubic lattice discretization, \emph{e.g.} for the solution of partial differential equations.

\begin{figure}
\includegraphics[scale=0.13]{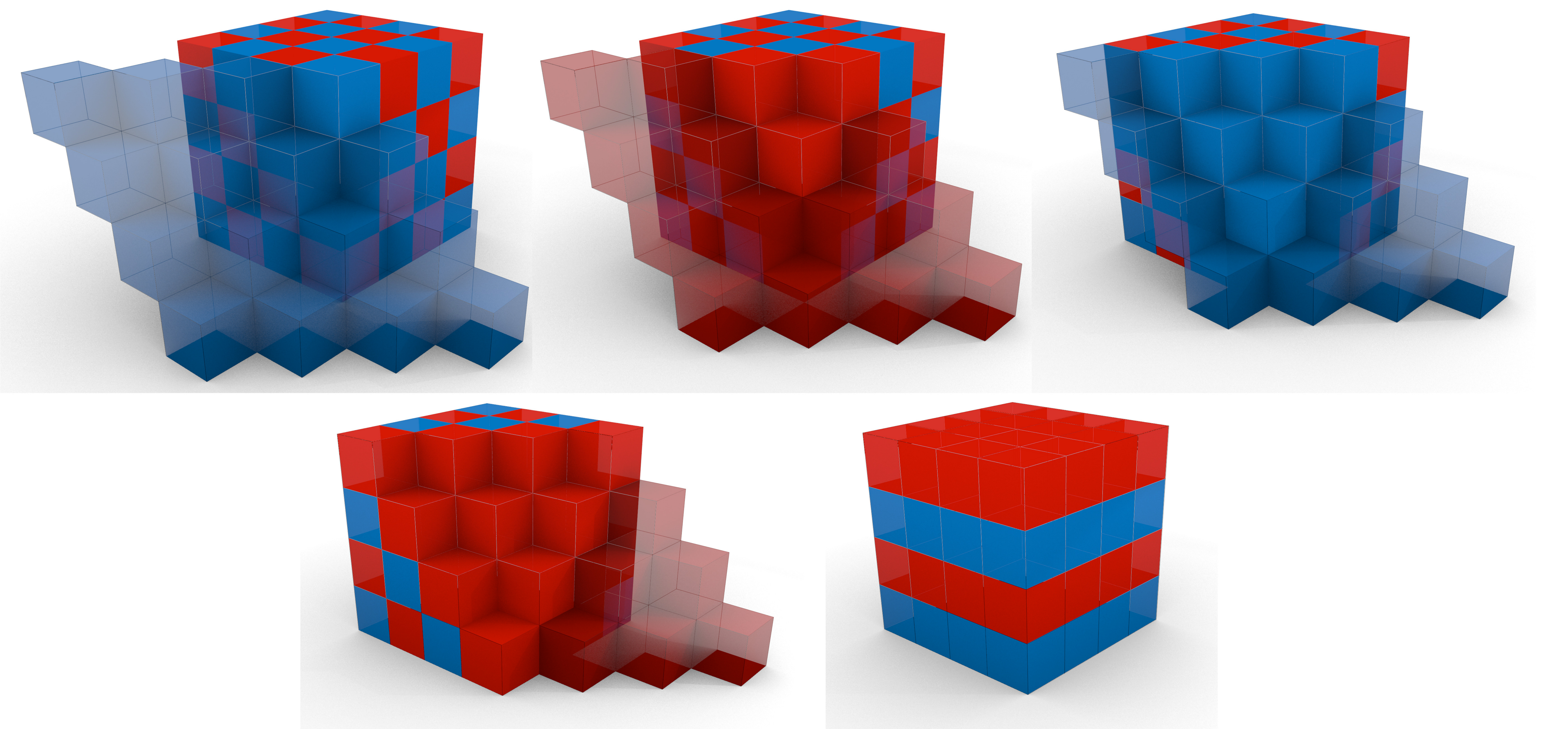}
\caption{\label{fig:CubeSlicing}A depiction of the slicing procedure. Lighter cells are the periodic ones.}
\end{figure}

\begin{figure}
\includegraphics[scale=0.75]{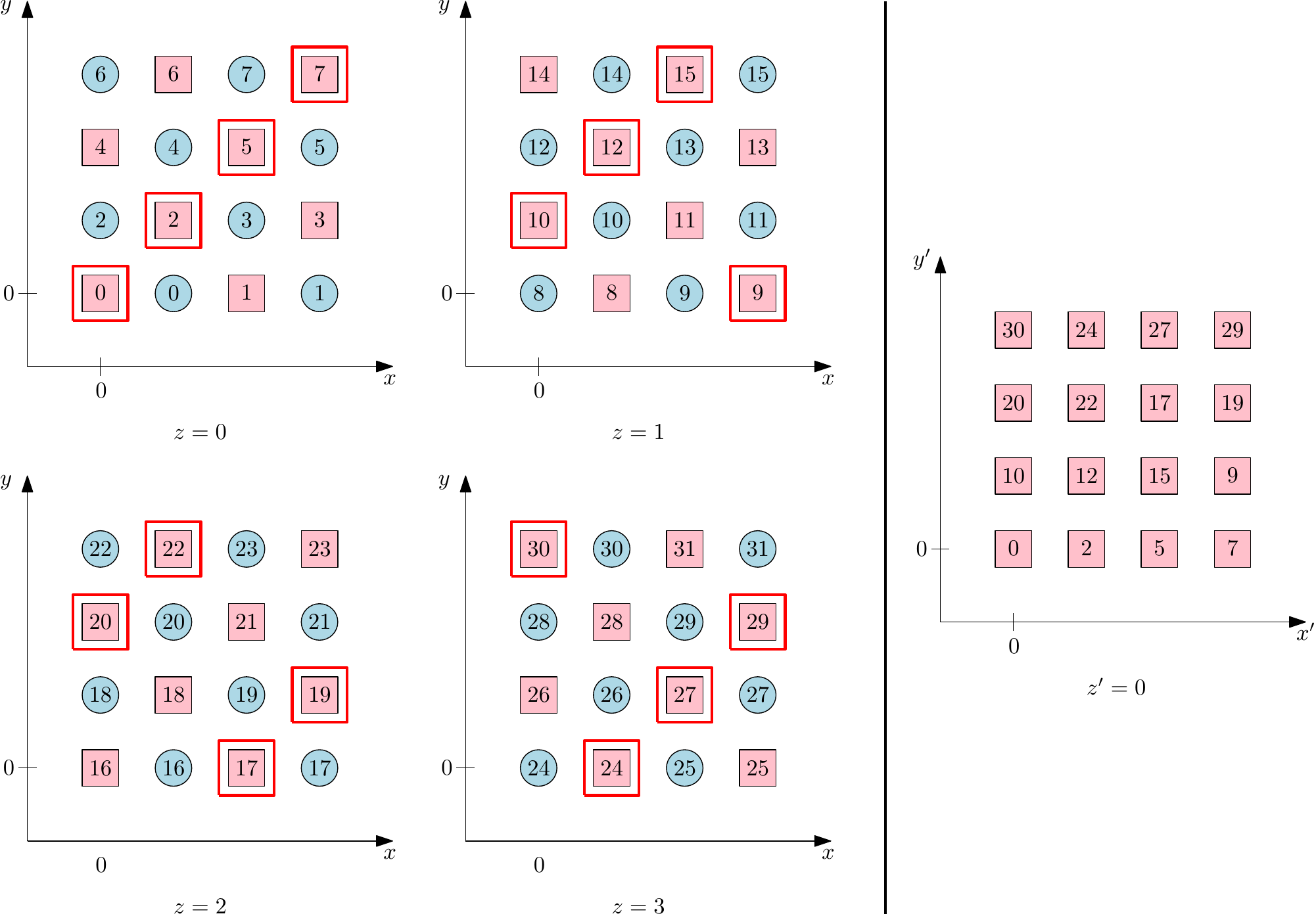}
\caption{\label{fig:SliceMapping}Mapping from the separated allocation checkerboard scheme to the sliced one for a lattice with $L=4$. All four planes of the real lattice are involved in the definition of the first plane. The procedure is iterated starting from the $z=1$ plane taking the blue diagonal: a blue slice is obtained for $z'=1$ and so on.}
\end{figure}

We now discuss the features of a new cubic stencil access pattern which we will refer to as \textbf{sliced}. We store red and blue spins in two separate arrays of Global Memory bound to two different textures. The novelty of the approach is in the spin arrangement. Here, we analyze the three-dimensional case, however the approach naturally extends to lower, \emph{i.e.} two-dimensional, and higher dimensional cases. In three dimensions, under the assumption of having periodic boundary conditions, there exist a way of separating the colours while keeping the cubic lattice topology: let us consider the cubic lattice starting from the origin of a Cartesian reference frame where the coordinates take on integer values, hence $\vec{x} \in \mathbb{Z}^3$; vertices belonging to planes orthogonal to the direction $\vec{n} = (1, -1, 1)$ are all one-coloured. That is the reason why we call this scheme \emph{sliced}. With periodic boundary conditions, vertices belonging to one slice will have all nearest neighbours either in the upper or in the lower slice. Such a slicing procedure is depicted in Fig. \ref{fig:CubeSlicing},
whereas in Fig. \ref{fig:SliceMapping} the transformation from the classic spin arrangement is shown.
Hence, one starts from a three-dimensional checkerboard and ends up with a cubic lattice where each horizontal plane is one-coloured.

It is easy to write down the transformation and its inverse given that the vertices coordinates read $\vec{x} = (x, y, z)$ and the transformed coordinates read $\vec{x}' = (x' , y' , z')$
\begin{equation}
\begin{cases}
x'= x\\
y'= y - x\\
z'= x - y + z
\end{cases}
\qquad
\begin{cases}
x = x'\\
y = y' + x'\\
z = z' + y'
\end{cases}
\end{equation}
recognizing in the equation for $z'$ the expression of a plane orthogonal to the direction $\vec{n} = (1, -1, 1)$. The generalization to any number of dimensions for the vector $\vec{n}$ is simply given by $n_i = (-1)^{i+1}$ so that in two dimensions one-coloured vertices lie on lines orthogonal to $\vec{n} = (1,-1)$, and in four dimensions they lie on hyperplanes orthogonal to $\vec{n} = (1,-1,1,-1)$ and so on. Hence, as long as one deals with regular (hyper)cubic lattices the scheme is completely general.

\begin{figure}
\includegraphics[scale=0.8]{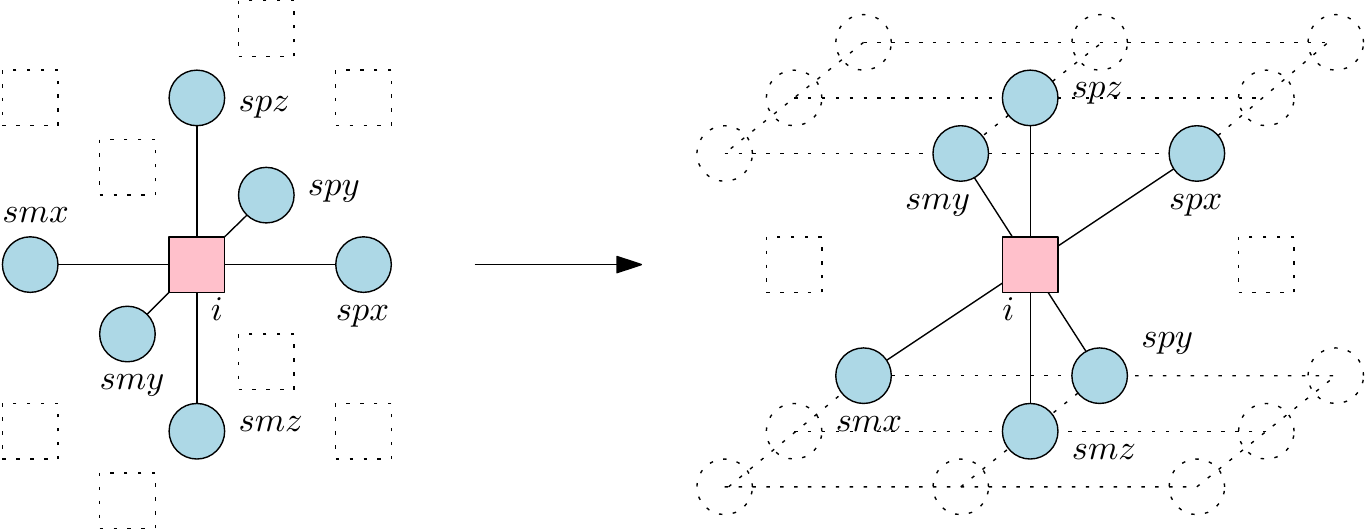}
\caption{\label{fig:CubicStencil}Cubic stencil for the real lattice (left) and for the transformed lattice (right).}
\end{figure}

The transformed coordinates of nearest neighbours are:
\begin{equation}
\begin{split}
\vec{x}_{spz} = (x, y, z + 1)\quad &\to \quad \vec{x'}_{spz} = (x', y', z' + 1)\\
\vec{x}_{smy} = (x, y - 1, z)\quad &\to \quad \vec{x'}_{smy} = (x', y' - 1, z' + 1)\\
\vec{x}_{spx} = (x + 1, y, z)\quad &\to \quad \vec{x'}_{spx} = (x' + 1, y' - 1, z' + 1)\\
&\\
\vec{x}_{smz} = (x, y, z - 1)\quad &\to \quad \vec{x'}_{smz} = (x', y', z' - 1)\\
\vec{x}_{spy} = (x, y + 1, z)\quad &\to \quad \vec{x'}_{spy} = (x', y' + 1, z' - 1)\\
\vec{x}_{smx} = (x - 1, y, z)\quad &\to \quad \vec{x'}_{smx} = (x' - 1, y' + 1, z' - 1)\\
\end{split}
\end{equation}
where the labels $spz,\ldots$ are self-explaining. The order of the new coordinates makes apparent that for the calculations of, say, $\vec{x}'_{spz}$ it is possible to reuse $\vec{x}'_{smy}$ and $\vec{x}'_{spx}$:
\begin{equation}
\vec{x}_{spx}' = \hat{i}'(x' + 1) + \vec{x}_{smy}' =  \hat{i}'(x' + 1) + \hat{j}'(y' - 1) + \vec{x}_{spz}',
\end{equation}
where $\hat{i}'$, $\hat{j}'$ and $\hat{k}'$ are the unit basis vector for the $x'$, $y'$ and $z'$ direction respectively. This feature gives some advantage in terms of calculations for the memory accesses, and it does not hold for the standard expressions.
Thus, the cubic stencil layout in both set of coordinates looks as in Fig. \ref{fig:CubicStencil},
and periodic boundary conditions apply also to the new coordinates.

Memory transactions for bulk spins are coalesced with some per-warp redundancy for $smy/spx$ and $spy/smx$  which is completely handled by the memory hierarchy. The new layout naturally shows a two-dimensional locality of data. Indeed, such a remapping is in the same spirit as the one proposed in \cite{Ferrero20121578}, with the difference that being this approach geometric it can be extended to other dimensionalities as shown above.

After having explained the memory arrangement we finally explain the spins-threads mappings. Some general remarks are in order:
\begin{itemize}
\item we simulate four replicas at once, systems with the same quenched disorder but different initial conditions and evolutions, which are stored in different arrays. Such a choice is a common practice \cite{Janus2013Eq}. However, the extension to an arbitrary number of replicas is trivial, requiring to handle a new stride in the kernels.
\item colours and couplings are bound to textures in order to delegate addresses calculations to the texture hardware rather than to CUDA cores. Indeed one colour is constant while updating the other.
\item for the sliced scheme, since the arrays are separated, one does not really need to consider $z'$ as running from 0 to $L-1$, but rather from 0 to $L/2 - 1$. By assigning the same $z'$ label to pairs of differently coloured $x'-y'$ planes and starting from the bottom with red vertices, the blue $spz$ of a red vertex $i$ has its same index, \emph{i.e.} $spz=i$, whereas the blue vertex $j$ will have its bottom red spin at $smz = j$. This scheme further simplifies calculations.
\item Currently, physically interesting behaviour of the EA3D model can be studied only for relatively small sizes, hence we should try to saturate the GPU resources also for small-size lattices. The easiest way to achieve this goal is to simulate different coded disorder realizations numbered by $k$. Thus, the stride separating in the spins arrays different coded samples is $L^3/2 = V/2$.

As far as we know only in \cite{Fang2014,MarcoBaity2013} such a technique has been adopted and indeed it is possible to sustain almost stable performances while varying the linear size $L$. We reserve the $y,z$ block grid dimension as disorder index, \emph{i.e.} different realizations of the couplings, which seems a reasonable choice since $\texttt{gridDim.y,z} < 65536$.

\item couplings are indexed as if they were red vertices and they are allocated in six different arrays \texttt{Jpx, Jpy, Jpz, Jmx, Jmy, Jmz}. This choice introduces an asymmetry in the kernels which can easily be fixed by allocating a copy of the couplings suitably transformed in order to be indexed as blue vertices. However, we do not show the results since the difference in the performances of the two updating kernels is negligible.
\end{itemize}

We implemented two different ways of mapping the vertices index to the threads index:
\begin{enumerate}
\item a one-dimensional mapping that associates $s$ vertices to a single thread. The value of $s$ can be tuned in order to find the best performance. Though, one needs to compute two divisions and two modulus operations in order to calculate nearest neighbours indices.
\item a multi-dimensional mapping exploiting the grid algebra provided by the GPU which allows to avoid divisions and modulus operations at the price of a more rigid choice for the total number of threads. The corresponding Kernels are tagged as \emph{Grid}.
\end{enumerate}

We tested the new access scheme comparing it with an implementation of the classic checkerboard spins arrangement, which we will refer to as \textbf{standard}, and with another scheme \cite{Bernaschi20111265} using mainly bitwise operations, which however works only in the case $L=2^{\ell}$. We will refer to this last implementation as \textbf{bitwise}.

We also wrote the Grid version of the standard scheme so that we end up with five different kinds of kernels: bitwise, standard, standard-Grid, sliced and sliced-Grid.

\section{Pseudo-Random Numbers Generators}
We chose to implement as a baseline the so-called Lehmer-Park-Miller MINSTD Linear Congruential PRNG which is defined as
\begin{equation}
R_{n+1} = (16807\, R_n) \mbox{mod} (2^{31} - 1),
\end{equation}
Its period is a prime number, more precisely a Mersenne prime $M_{31} = 2^{31} -1$. This generator can be used for the coupling values $J_{ik}$ and it is also a reasonable choice for the critical off-equilibrium relaxation dynamics under the hypothesis of a number of instances no larger than the period.

One difficulty comes with the implementation of the module which cannot be carried out by hardware truncation, so that we need to directly handle the overflow due to the multiplication and then take the module. This can be done by means of a swap 64-bit variable or by means of 32-bit variables only as proposed by Carta in \cite{Carta1990, MINSTDSite}. The latter solution does not require the module operation. We followed the implementation proposed in \cite{MINSTDSite} but we substituted the conditional statements with bitwise operations to avoid warp branchings.

However, since we plan to extend this MC implementation to the equilibrium regime we also developed a GPU version of the Parisi-Rapuano PRNG \cite{ParisiRapuano} which is mostly used in the spin-glass community. It is a lagged-Fibonacci-like PRNG with a minimal state of 62 words. One instance of the generator reads
\begin{equation*}
\begin{split}
&\texttt{ira[i] = ira[i - 24] + ira[i - 55];}\\
&\texttt{R = ira[i]\^\,ira[i - 61];}\\
\end{split}
\end{equation*}
where \texttt{ira} denotes the state array and \texttt{R} is the new random number. A
common approach \cite{Weigel_PerformancePotential, WeigelPRNG2012} consists in exploiting the lags and let the threads in a block share one or more states which can be concurrently updated storing them in Shared Memory. However lags as those of the Parisi-Rapuano PRNG are not well-suited for this scheme \cite{WeigelPRNG2012}.

Hence, we propose a new simple alternative: allocate an array of $N_{threads}\times N_{state}$ entries and let each thread access it with its own global grid index and load the lagged entries just using a stride, \emph{i.e.} the number of threads. Thus, defining \texttt{d\_threads} as the number of threads and \texttt{globalId} as the thread global grid index, a sketch of the kernel implementation simply reads
\begin{equation*}
\begin{split}
\texttt{swap = ira[(i - 24)*d\_threads + globalId] }&\\
\texttt{     + ira[(i - 55)*d\_threads + globalId];}&\\
\texttt{R = swap\^\,ira[(i - 61)*d\_threads + blobalId];}&\\
\texttt{ira[i*d\_threads + globalId] = swap;}&\\
\end{split}
\end{equation*}
although in an actual implementation one has to take into account the periodic conditions for the access to the state\footnote{More details can be found in the appendix}. 

In order to show the validity of the new scheme we chose to implement the widely known Mersenne Twister MT19937 and compare its performance to that of the cuRand MTGP which is a modified version of the Mersenne Twister. For the host API comparison we use the criterion proposed in \cite{PRAND}. Nonetheless, we propose as standard benchmark for a PRNG its kernel version counting the fraction of odd numbers (just as the example reported in the cuRand manual \cite{cuRandDeviceApiSite}). Such a benchmark should be more suitable for kernel-use PRNGs.

Results are reported in Table \ref{tab:PRAND} for the PRAND test, and in Table \ref{tab:MYTEST} for the device API test. Tests were run on GTX 680, GTX Titan, Tesla M2090 and Tesla K20x GPUs. Looking at Table \ref{tab:PRAND}, where we report the execution times for filling an array of $2^{29}$ single-precision floating point variables, we see that our implementation of MT19937 runs roughly twice as fast as the cuRAND MTGP32 implementation. We could only test the most recent host API cuRAND implementation of the MT19937 on the GTX Titan and not on the K20x (the M2090 is ruled out being too old), and our implementation performs 44\% slower than cuRAND. The large $t_{\scriptscriptstyle{INIT}}$ values for our implementation are due to the fact that the seed are read from the system random pool, slowing down the process. It is clearly possible to reduce those times implementing some initialization algorithms as those proposed in \cite{MT19937Site}. In Table \ref{tab:MYTEST} the metric is changed to the number of PRNG instances per second. The trends are qualitatively the same although our implementation of the MT19937 on the GTX Titan runs almost three times faster than the cuRAND MTGP32.

However, such benchmarks only give an indication of the speed of different PRNGs. As we will see one should always compare different PRNGs in a real-life application.

\begin{table}
  \begin{tabular}{ l | c | c | c || c | c | c || c | c | c |}
 \toprule
   & \multicolumn{3}{ c ||}{Tesla M2090} & \multicolumn{3}{ c ||}{Tesla K20X} & \multicolumn{3}{ c |}{GTX Titan} \\
 \hline
  PRNG & $t_{\scriptscriptstyle{INIT}}$(s) & $t_{\scriptscriptstyle{GEN}}$(s) & $t_{\scriptscriptstyle{TOT}}$(s) & $t_{\scriptscriptstyle{INIT}}$(s) & $t_{\scriptscriptstyle{GEN}}$(s) & $t_{\scriptscriptstyle{TOT}}$(s) & $t_{\scriptscriptstyle{INIT}}$(s) & $t_{\scriptscriptstyle{GEN}}$(s) & $t_{\scriptscriptstyle{TOT}}$(s)\\
 \hline
\hline
cuRand MTGP32 & 0.09 & 12.34 & 12.43 & 0.12 & 13.46 & 13.58 & 0.21 & 10.11 & 10.32\\
\hline
cuRand XORWOW & 0.01 & 2.91 & 2.92 & 0.01 & 2.90 & 2.91 & 0.01 & 2.31 & 2.32\\
\hline
cuRand MT19937& 0 & 0 & 0 & 0 & 0 & 0 & 0.01 & 3.23 & 3.24\\
\hline
\hline
MT19937       & 3.86 & 6.40 & 10.26 & 4.63 & 6.12 & 10.75 & 3.92 & 4.66 & 8.58\\
\hline
Parisi-Rapuano& 0.40 & 8.17 & 8.57 & 0.45 & 5.87 & 6.32 & 0.41 & 4.18 & 4.59\\
\hline
MINSTD        & 0.01 & 1.72 & 1.73 & 0.01 & 1.34 & 1.35 & 0.01 & 1.13 & 1.14 \\
\hline
\end{tabular}
\caption{PRAND benchmark \cite{PRAND} results using cuRand host API. The task consists in filling an array of $2^{29}$ single-precision floating point variables. In the upper half of the Table, cuRand library results are reported whereas in the lower half those of our implementations. Two different measures are reported: $t_{\scriptscriptstyle{INIT}}$ is the time needed to initialize the PRNG; $t_{\scriptscriptstyle{GEN}}$ is the generation time. For the M2090 ECC is off, while for K20x ECC is on.}
\label{tab:PRAND}
\end{table}

\begin{table}
\begin{tabular}{ l | c | c | c | c |}
\toprule
PRNG & M2090 & K20X & GTX Titan & GTX 680\\
\hline
cuRand MTGP32  & $4.5 \cdot 10^{9}$  & $3.9 \cdot 10^{9}$ & $5.2 \cdot 10^{9}$   & $5.1 \cdot 10^{9}$\\
\hline
cuRand XORWOW  & $2.9 \cdot 10^{10}$ & $7.6 \cdot 10^{10}$ & $10.7 \cdot 10^{10}$ & $6.1 \cdot 10^{10}$\\
\hline
\hline
MT19937        & $10.1 \cdot 10^{9}$ & $10.7 \cdot 10^{9}$ & $14.1 \cdot 10^{9}$  & $9.6 \cdot 10^{9}$\\
\hline
Parisi-Rapuano & $9.4 \cdot 10^{9}$  & $12.5 \cdot 10^{9}$ & $16.8 \cdot 10^{9}$  & $8.3 \cdot 10^{9}$\\
\hline
MINSTD         & $4.1 \cdot 10^{10}$ & $7.6 \cdot 10^{10}$ & $8.9 \cdot 10^{10}$  & $6.7 \cdot 10^{10}$\\
\hline
\end{tabular}
\caption{Device API test. Number of instances per second. The launch configuration is the following: 64 blocks of 256 threads, each thread producing $2^{15}$ instances, repeated 10 times. For the M2090 ECC is off, while for K20x ECC is on.}
\label{tab:MYTEST}
\end{table}
\section{Asynchronous Multispin-Coding}
Multispin-coding techniques are rooted in lattice gauge theory simulations \cite{PhysRevLett.42.1390_MSC3}. They have been employed later in Ising models simulations \cite{MSC0, PhysRevB.33.7841_MSC1, PhysRevB.35.5382_MSC2, Ito:1990:MCS:110382.110607}. The search for a close packing of data was motivated by the limited memory resources of that time, and by the intrinsic bit-level parallelism which can be obtained through bitwise operations. Indeed, since the quantities involved in the simulations are two-valued, \emph{i.e.} $\sigma_i \in \{-1,+1\}$ and $J_{ik} \in \{-1,+1\}$, the optimal solution is to store couplings and spins in single bits rather than use a single byte, \emph{e.g.} using a \texttt{char}.

Multispin-coding comes in two different flavours:
\begin{itemize}
\item synchronous multispin-coding (SMSC) consisting in storing in one word spins belonging to one single system, usually aligned along one specific direction. This allows to get faster simulations in terms of wall-clock time compared to a simple
one-variable-one-spin setting. Indeed, such a technique is used in the Janus supercomputer \cite{JanusTempi} for reaching thermal equilibrium. Clearly, the update of each bit-spin requires one instance of the PRNG;

\item asynchronous multispin-coding (AMSC) consists in storing spins belonging to different systems, located at the same vertex, in the same word. The total wall-clock time does not decrease, but it is possible to update all spins contained in a word with only one instance of the PRNG at the cost of the introduction of a certain amount of correlation, which can be taken care of easily.
\end{itemize}

We chose to implement the AMSC because we were interested in the off-equilibrium critical relaxation regime, hence being able to simulate a large number of samples is preferable over obtaining a long simulation time. The AMSC for spin systems was clearly explained in \cite{Ito:1990:MCS:110382.110607} where each system was considered to be at a different temperature. We are aware of AMSC implementations on GPU: \cite{Block20101549} for the 2D Ising model and \cite{Fang2014} for the EA3D model with external field. In particular in \cite{Fang2014} the proposed AMSC techinque stores in one word spins of the same system, \emph{i.e.} with the same couplings, which are evolved at different temperatures. This scheme has been adopted for implementing the PT dynamics. Transition probabilities are stored in a look-up table indexed by the energy difference $\Delta E$ of the proposed flip and the spin direction (with respect to an external magnetic field). Hence, the swap of two temperature-replicas simply requires to swap two lines in the look-up table. However, in order to speed up the access to the look-up table, some space in the spin words is reserved so that not all bits of a word codify for a spin. We will see that for non-PT dynamics this represents a bottle-neck for memory use efficiency. Again, each spin update is served by one PRNG instance.

As we anticipated, we associate to each spin a different disorder realization thus only one PRNG instance is needed for all spins contained in a word. Considering the contribution to the Hamiltonian due to a single cubic stencil, it is clear that the possible energy differences after a proposed spin flip on $\sigma_a$ are
\begin{equation}
\Delta E = H[\{\sigma_{i\neq a}, -\sigma_a\}] - H[\{\sigma_{i\neq a}, \sigma_a\}] = -12, - 8, -4, 0, 4, 8, 12.
\end{equation}
The Metropolis dynamics is defined by the acceptance probability
\begin{equation}
P_{\mbox{flip}}(\Delta E) =
\begin{cases}
1,&\quad \Delta E \leq 0\\
e^{-\beta \Delta E},& \quad \Delta E > 0
\end{cases}
\end{equation}
where $\beta = T^{-1}$ is the inverse temperature. The value of $P_{\mbox{flip}}(\Delta E > 0)$ has to be compared to a flat-distributed random number $r\in[0,1]$ so that if $r < P_{\mbox{flip}}(\Delta E > 0)$ the proposed flip is accepted otherwise it is rejected. However, since PRNGs are defined for integers, one does not really need to use a normalized $r$. The most direct way is to multiply the transition probability for the value of the biggest random number $R_{max}$ and compare it with the PRNG instance $R$, \emph{i.e.} $R \lessgtr R_{max}\exp(-\beta \Delta E)$. We label the non-trivial normalized transition probabilities as $R_{max}\exp(-\beta \Delta E)=\texttt{EXP12}, \texttt{EXP8}, \texttt{EXP4}$. We employ the following mapping of spins and couplings to bits
\begin{equation}
\begin{split}
J_{ik}=-1\;\to\;\texttt{J}_{ik}\texttt{=1},\qquad\sigma_i=-1\;\to\;\texttt{s}_i\texttt{=0},\\
J_{ik}=+1\;\to\;\texttt{J}_{ik}\texttt{=0},\qquad\sigma_i=+1\;\to\;\texttt{s}_i\texttt{=1}.
\end{split}
\end{equation}
The value of the interaction energy with one of the nearest neighbours is then converted for each bit as
\begin{equation}
\begin{split}
-J_{ik}\, \sigma_i\, \sigma_k = -1\;\to\;\texttt{e}_{ik} = \texttt{J}_{ik}\mbox{\texttt{\^}}\texttt{s}_i\mbox{\texttt{\^}}\texttt{s}_k = \texttt{0},\\
-J_{ik}\, \sigma_i\, \sigma_k = +1\;\to\;\texttt{e}_{ik} = \texttt{J}_{ik}\mbox{\texttt{\^}}\texttt{s}_i\mbox{\texttt{\^}}\texttt{s}_k = \texttt{1}.\\
\end{split}
\end{equation}
If we sum the six energy variables per stencil $\texttt{e}_{ik}$ we obtain a three-bit result
\begin{equation}
\sum_{k} \texttt{e}_{ik} = (\texttt{sum2, sum1, sum0}) = 2^2\times\texttt{sum2} + 2\times\texttt{sum1} + \texttt{sum0},
\end{equation}
which directly maps to the seven possible values of $\Delta E$, since flipping the spin leads to flip the partial values $\texttt{e}_{ik}$.
\begin{equation}
\begin{split}
&(\texttt{0, 0, 0})=0\quad \to \quad \Delta E = -12\\
&(\texttt{0, 0, 1})=1\quad \to \quad \Delta E = -8\\
&(\texttt{0, 1, 0})=2\quad \to \quad \Delta E = -4\\
&(\texttt{0, 1, 1})=3\quad \to \quad \Delta E = 0\\
\end{split}
\qquad
\begin{split}
&(\texttt{1, 0, 0})=4\quad \to \quad \Delta E = 4\\
&(\texttt{1, 0, 1})=5\quad \to \quad \Delta E = 8\\
&(\texttt{1, 1, 0})=6\quad \to \quad \Delta E = 12\\
&\\
\end{split}
\end{equation}

Now, the aim is to define a mask in order to flip the right spins with a XOR operation
\begin{equation}
\texttt{spin} = \texttt{spin}\; \hat{} \;\texttt{mask};
\end{equation}
As a first step we compare the random number \texttt{R} with the non-trivial transition probabilities defining the variables
\begin{equation}
\begin{split}
&\texttt{cond12 = -(R < EXP12);}\\
&\texttt{cond8 = -(R < EXP8);}\\
&\texttt{cond4 = -(R < EXP4);}\\
\end{split}
\end{equation}
\emph{i.e.} if $\texttt{R} < \texttt{EXP4}$ then \texttt{cond4 = 0xffffffff} (all bits set equal to one), whereas if $\texttt{R} > \texttt{EXP4}$ then \texttt{cond4 = 0x00000000} (all bits equal to zero). Clearly, if \texttt{cond12 = 0xffffffff}, \emph{i.e.} the most improbable flip can be accepted, then all spins must be flipped. Also all spins with \texttt{sum2 = 0} must be flipped so that we can write
\begin{equation}
  \texttt{mask = cond12 | (} \mbox{\texttt{\~}}\texttt{sum2)};
\end{equation}
where the \texttt{|} stands for the bitwise OR operator. We still need to handle the two remaining non-trivial cases corresponding to $\Delta E = 4,~8$. A first selection is obtained by using \texttt{sum2} as a mask, although we must discard the case $\Delta E = 12$, hence we write \texttt{sum2 \& (sum2 \^}\texttt{ sum1)}. The flipping condition for the cases $\Delta E = 4, 8$, when $\texttt{R} < \texttt{EXP8}$, simply reads \texttt{(sum2 \& (sum2 \^}\texttt{ sum1)) \& cond8}. The last step is to consider $\Delta E = 4$ when $\texttt{R} < \texttt{EXP4}$ which leads to \texttt{(sum2 \& (sum2 \^}\texttt{ sum1)) \& (cond8 | (cond4 \& (\~}\texttt{sum0))}. All in all the mask reads
\begin{equation}
\begin{split}
  \texttt{mask =}&\texttt{ cond12 | (} \mbox{\texttt{\~}}\texttt{sum2)}\\
&\texttt{| ((sum2 \& (sum2}\mbox{\texttt{\^}} \texttt{sum1)) \& (cond8 | (cond4 \& (}\mbox{\texttt{\~}}\texttt{sum0))))};
\end{split}
\end{equation}
This expression has the same number of bitwise operations of the natural extension of \cite{Ito:1990:MCS:110382.110607}.

\subsection{Results}
We present now the results concerning the performances of the different GPU implementations which are labeled as \textbf{sliced}, \textbf{standard} and \textbf{bitwise}. The \textbf{sliced} one uses the sliced checkerboard scheme we propose in this work, whereas the \textbf{standard} and \textbf{bitwise} implementations are based on the usual checkerboard scheme with the difference that the last one only works for linear sizes which are powers of two, $L=2^{\ell}$ and the calculations are implemented mainly through bitwise operations. We checked that all these schemes give the same bit-to-bit results so that they are completely equivalent\footnote{Precisely, the bitwise check between the sliced and the other implementations requires to remap all random numbers after one of the two colours has been updated.}.

Before discussing the results let us define the principal metric we will use in order to measure performances: the pico-second-spin-flip $\mbox{psFlip}_{n,\mbox{x}}$ that is how many pico-seconds are needed in order to reject or accept a proposed spin-flip. Here $n$ stands for the number of GPUs and `x' for the used PRNG. The mathematical definition is the following
\begin{equation}
\mbox{psFlip}_{n,\mbox{x}}(L, k) = t_{\mbox{sw}} \cdot n \cdot \left( 32 \cdot k \cdot 4 \cdot L^3 \right)^{-1},
\label{eq:psFlip}
\end{equation}
where $32\cdot k$ is the number of different disorder realizations (32 multispin-coded times $k$ different codings), 4 is the number of simulated replicas, and $t_{\mbox{sw}}$ is the wall-clock time needed to perform one sweep, \emph{i.e.} update red and blue spins, for \emph{all} disorder realizations: $t_{\mbox{sw}}$ is always measured on a single node. Data were taken for four different GPUs: GTX 680, GTX Titan, Tesla M2090 and Tesla K20x.

Let us begin by studying a problem that has been hardly explored in the past: how to saturate the GPU resources for small lattices. The solution we adopted, as others did \cite{Fang2014, MarcoBaity2013}, is to allocate at the same time different systems.
\begin{figure}
\includegraphics[scale=1.0]{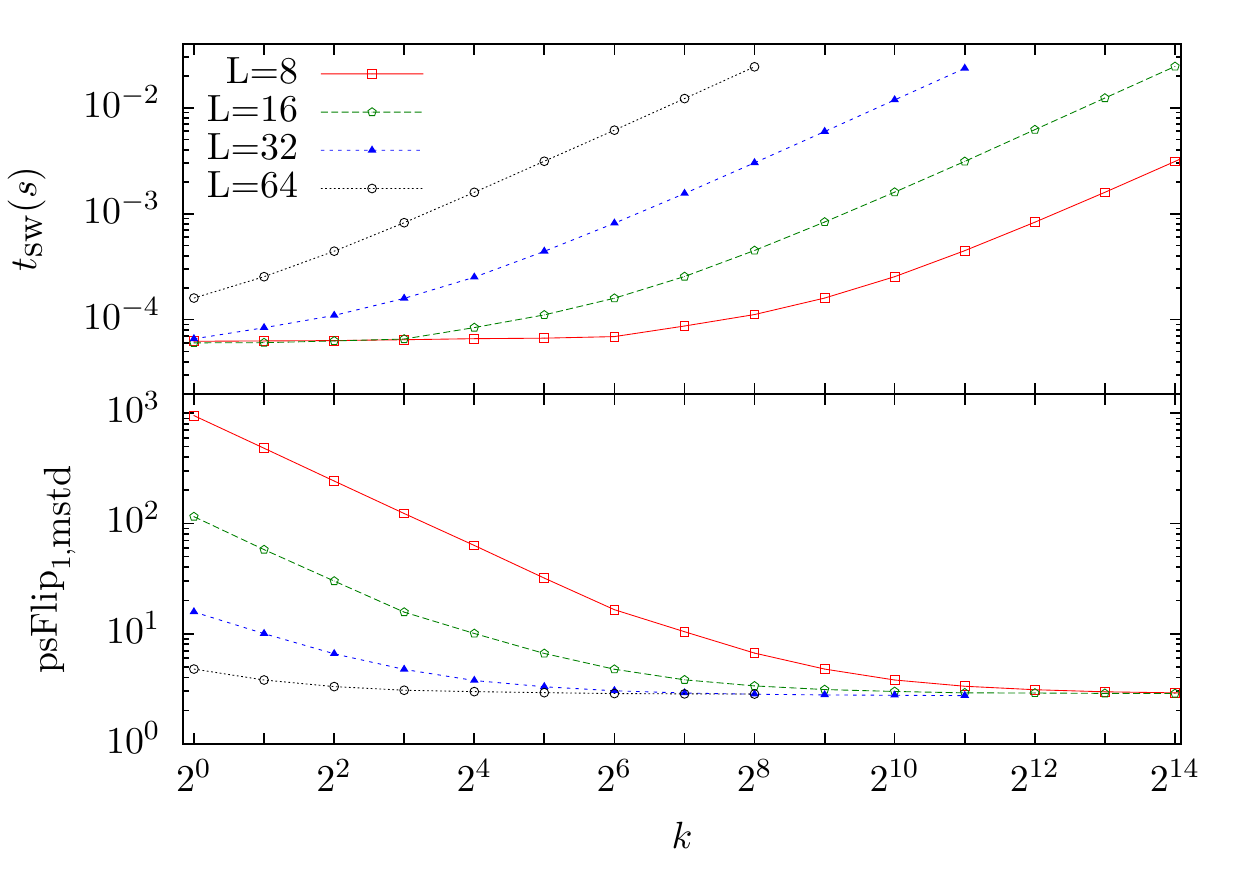}
\caption{Values for $t_{\mbox{sw}}$ and $\mbox{psFlip}_{1,\mbox{mstd}}$, \emph{i.e.} the number of pico-seconds needed to accept or reject a flip proposal with the MINSTD PRNG, as a function of the number of coded systems $k$ for a GTX Titan. Data refer to the best performances at varying grid launch parameters for a given value of $k$ for different sizes $L$.}
\label{fig:realTimes}
\end{figure}
As it is shown in Fig.\ref{fig:realTimes}, in the case of $L=8$, $t_{\mbox{sw}}$ is almost constant up to $k=64$ which means that simulating one or $64$ coded systems has the same cost for the GPU. This means that a factor $64$ can be gained for free. Indeed, this observation is important since the accessible physics for the EA3D is still confined to relatively small lattices, hence obtaining the best result also for $L\leq 32$ is crucial. We notice that even though, for $k>64$, $t_{\mbox{sw}}$ starts to increase, a linear regime is attained only for $k\geq 4096$ in the case $L=8$. Indeed, the case $L=32$ saturates the GPU almost at the beginning and the metric $\mbox{psFlip}_{1,\mbox{mstd}}$ only evolves from 4 psFlip to 3 psFlip, which however is a $\sim 25\%$ gain.

Now, in order to make a fair comparison with Janus FPGA hardware \cite{Janus2012Tech,Janus2_2014}, it is important to stress that those machines sustain comparable performances in terms of pico-seconds-spin-flip (16 psFlip for Janus and 3-5 psFlip for Janus II) for a \textbf{single} sample also for small lattice sizes. In our case we need to simulate several samples to saturate the GPU resources. Hence, Janus and Janus II are the fastest solution in terms of wall-clock time to bring a single sample to equilibrium and for small lattice sizes GPUs are still far away. A direct comparison with Janus supercomputers can be only performed when a single system is large enough to saturate the GPU resources. However, the game is subtle since saturation is attained only for large sizes which might be out of the domain of physical interest, at least for equilibrium simulations.

\begin{figure}[h!]
\centering
\subfigure{
\includegraphics[scale=1.0]{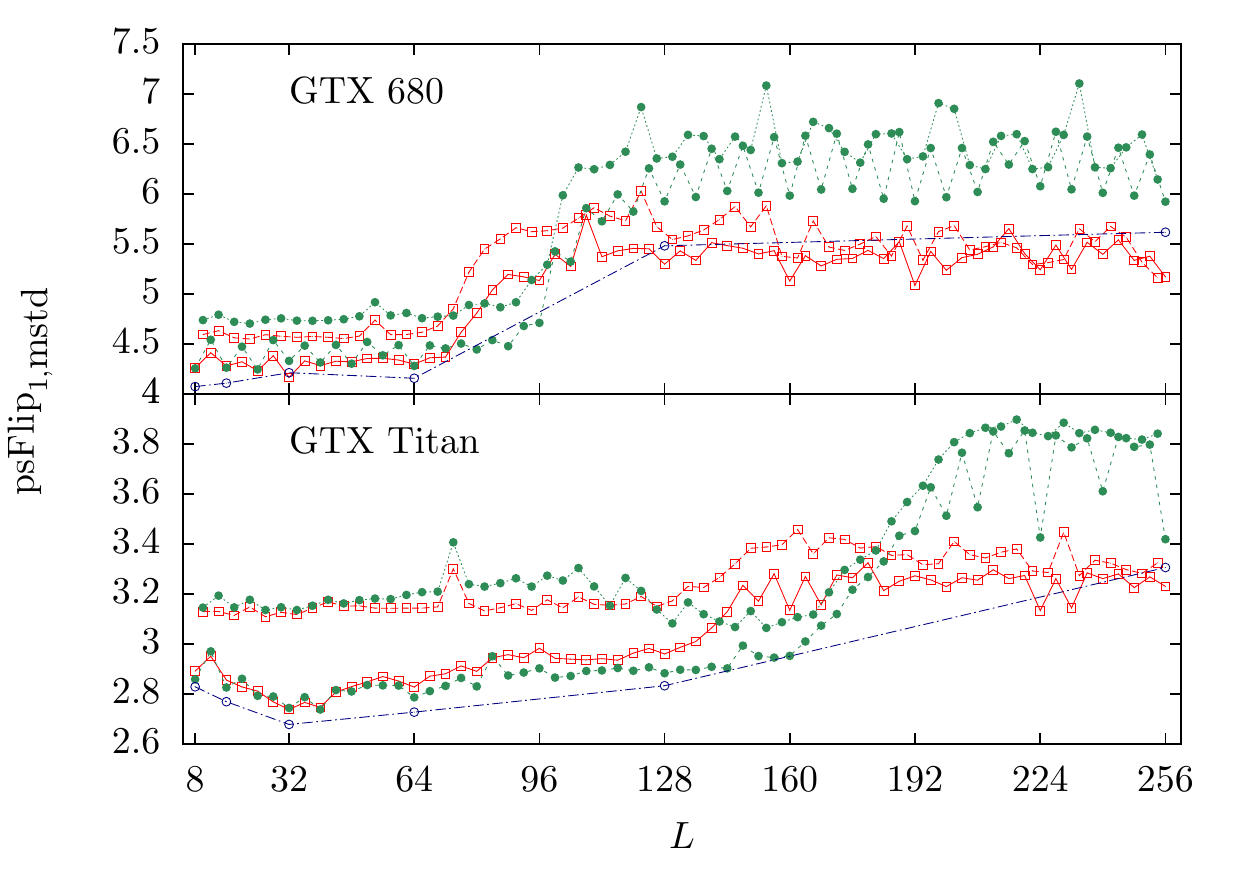}
\label{fig:psFlipGTX}
}
\subfigure{
\includegraphics[scale=1.0]{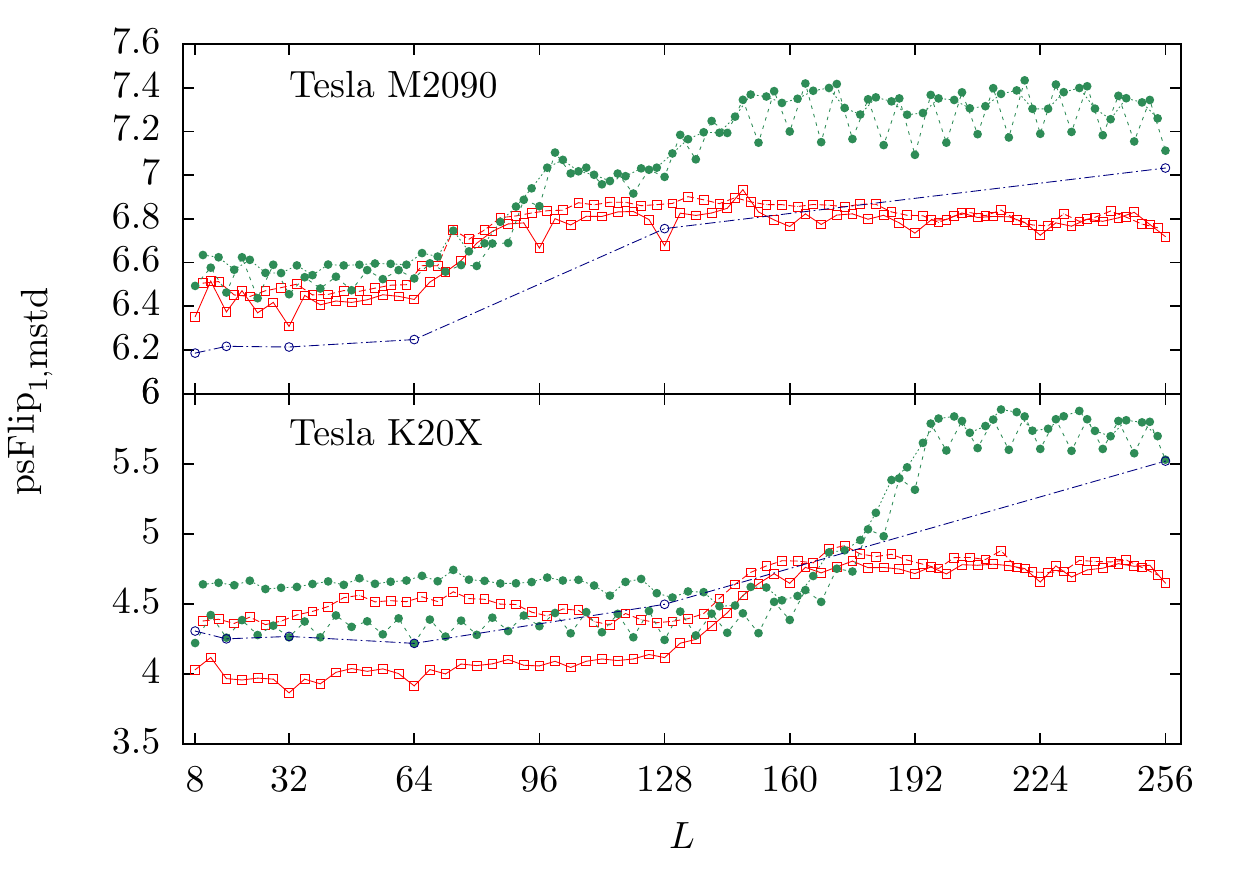}
\label{fig:psFlipTesla}
}
\caption{Best performances for $\mbox{psFlip}_{1,\mbox{mstd}}$. The red empty squares refer to the \textbf{sliced} implementation, the light-green filled circles refer to the \textbf{standard} implementation whereas the blue empty circles to the \textbf{bitwise} one. The value of $L_{thr}$ is larger for the GTX Titan and the Tesla K20x.}
\end{figure}

\begin{figure}[h!]
\centering
\subfigure{
\includegraphics[scale=1.0]{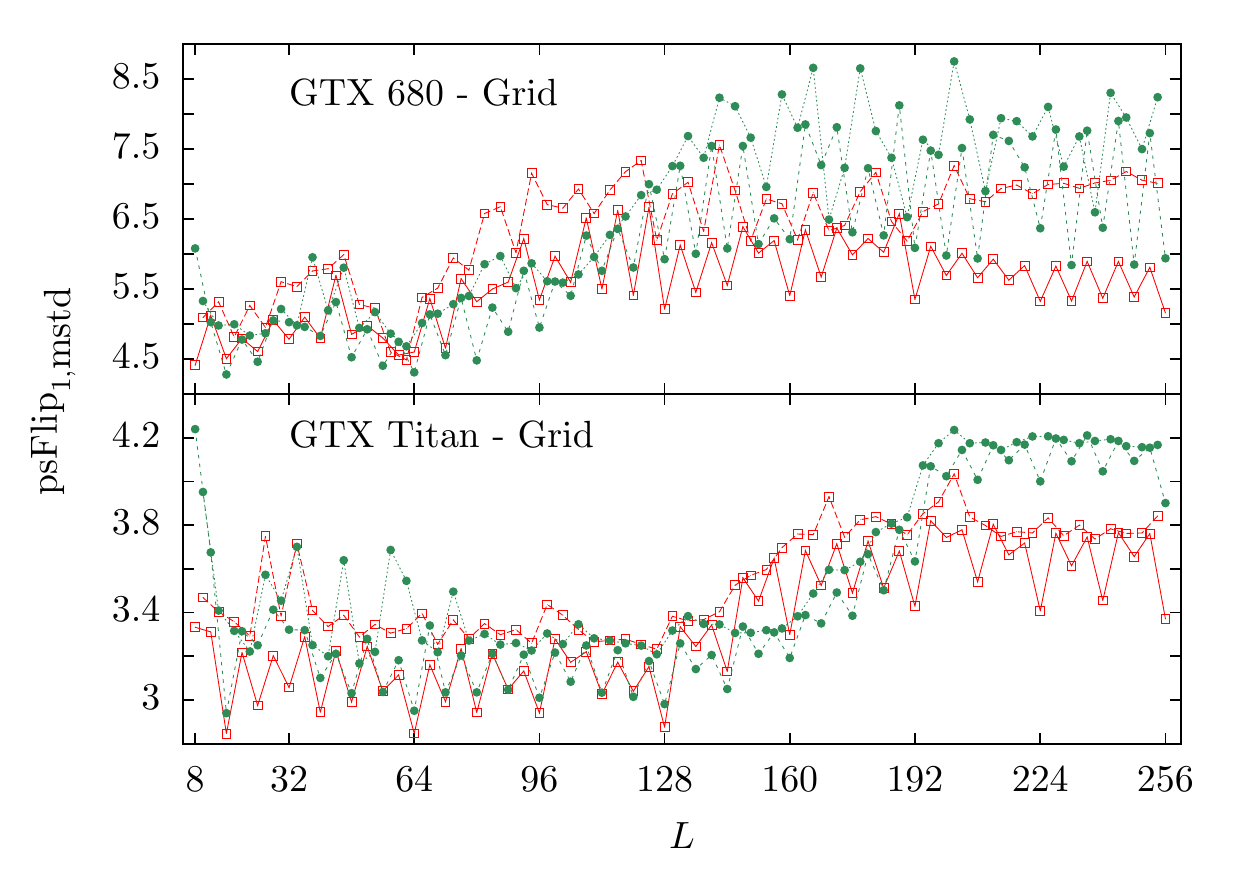}
\label{fig:psFlipGTXGrid}
}
\subfigure{
\includegraphics[scale=1.0]{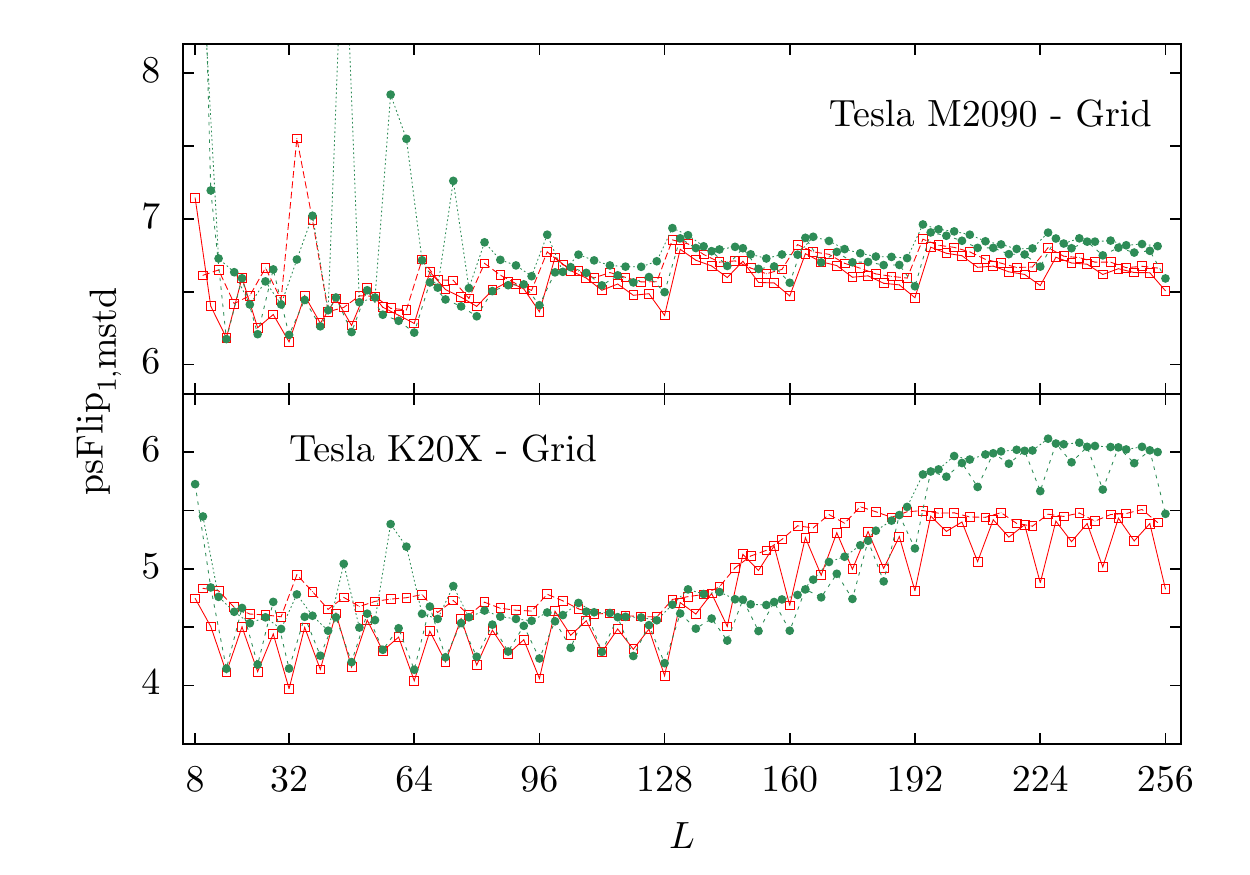}
\label{fig:psFlipTeslaGrid}
}
\caption{Best performances for $\mbox{psFlip}_{1,\mbox{mstd}}$ for Grid implementations. The red empty squares refer to the \textbf{sliced-Grid} implementation and the light-green filled circles refer to the \textbf{standard-Grid} implemention.}
\end{figure}

In Figures \ref{fig:psFlipGTX}, \ref{fig:psFlipTesla}, \ref{fig:psFlipGTXGrid} and \ref{fig:psFlipTeslaGrid} we report benchmarks results for different GPUs and different algorithms on all even lattice size in the range $8\leq L\leq 256$ . They all share the MINSTD as PRNG. Benchmarks were performed measuring the sweep wall-clock time $t_{\mbox{sw}}$ while varying $L$, $k$ and the grid configuration for the kernel, in order to find the best configuration for each lattice size, \emph{i.e.} only the best configurations times are reported. There are some qualitative features which are shared by the different GPUs
\begin{itemize}
\item the best performances are obtained in the first range of lattice sizes $L<L_{thr}$, where the threshold $L_{thr}$ varies according to the GPU and the algorithm, assuming larger values for latest GPUs; $L_{thr}$ is defined as the first value of $L$ for which $\mbox{psFlip}_{1,\mbox{mstd}}$ begins to grow significantly;
\item the sliced scheme performances get worse always before those of the standard scheme do;
\item the sliced scheme gives always the best performance for $L>L_{thr}$;
\item we split the data in two different branches defined by two subsequences of the lattice size $L_0 = 4m$ (faster) and $L_1 = 2(2m + 1)$ (slower), which converge for $L>L_{thr}$, and this splitting is most evident for small lattice sizes.
\end{itemize}

The sliced scheme worsen before the standard does probably because the latter deals with boundary conditions on the $y$-axis only after $L^2/2$ elements have been processed whereas for the sliced scheme the boundary conditions on the $y'$-axis are treated after $L^2$ elements. Hence, a cache hit for the standard scheme is more likely. It appears that the behaviour of the GPU memory is somehow correlated to the number of memory requests for the periodic boundaries. As a matter of fact, the following scaling relation $L_{thr}^{standard} \sim \sqrt{2}\,L_{thr}^{sliced}$ roughly holds.

Data related to the standard-Grid and sliced-Grid implementations are clearly less stable.We notice that for $L\gtrsim 8$ the standard-Grid implementation performs much worse than the sliced-Grid. This should be related to the fact that the block size is fixed to the number of one-coloured spins in a $z$ slice, which means for the standard-Grid scheme $L^2/2$, starting from 32 threads, and for the sliced-Grid $L^2$, starting from 64 threads. Clearly, having blocks which coincide with a warp is not an optimal choice for the GPU. Looking at the data for the Tesla M2090 in \ref{fig:psFlipTeslaGrid} there is a modulation as a function of the lattice size with a period $\Delta L = 32$. We notice that for such values of $L$ the blocks are always multiple of a warp.

The `Grid' algorithms perform slower than the others in the examined range so that we can safely discard this implementation choice which relies on the inherent algebra of the thread-grid indices. Hence, we will focus hereafter mainly on the non-grid implementations.
\begin{figure}[h!]
\includegraphics[scale=1.0]{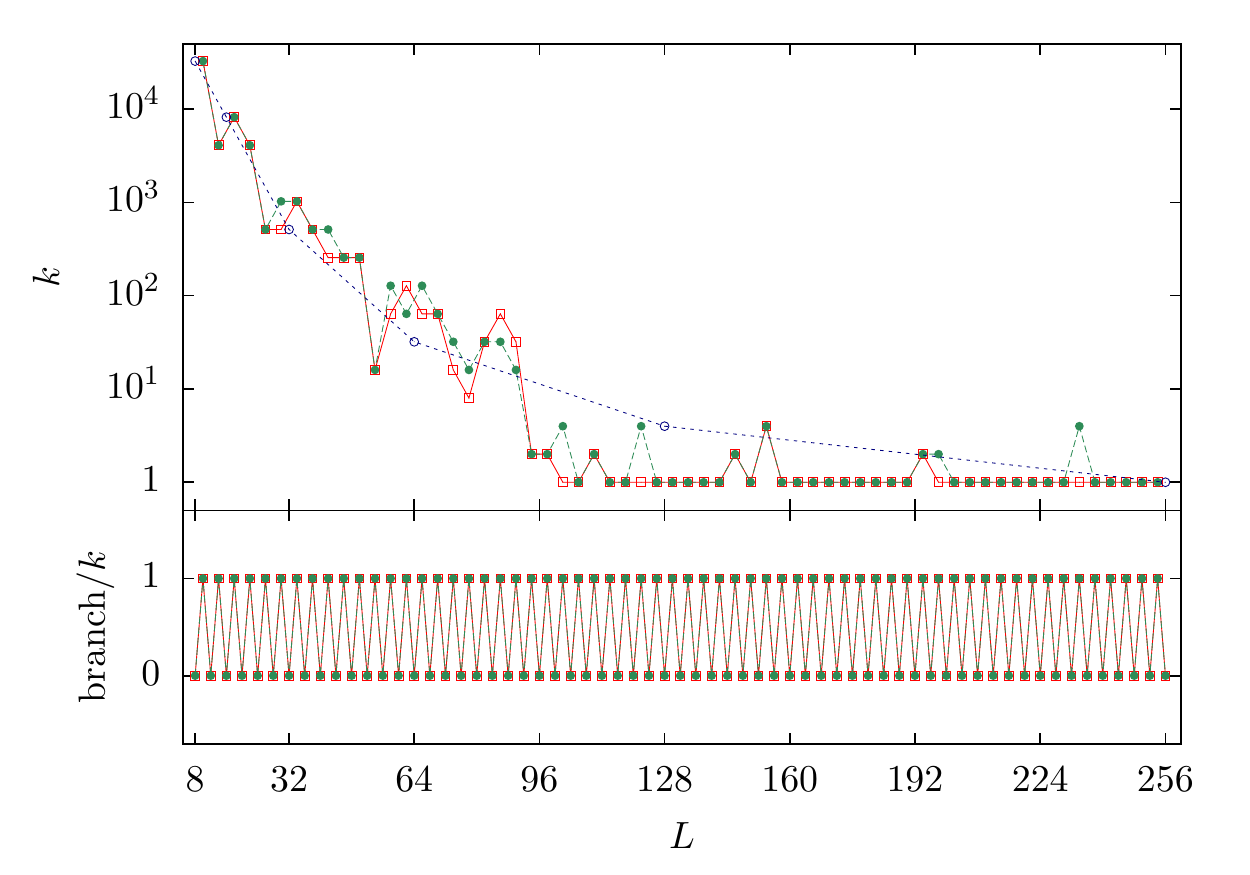}
\caption{Upper panel: best values of $k$ for the GTX 680. The red empty squares refer to the \textbf{sliced} implementation, the light-green filled circles refer to the \textbf{standard} implementation whereas the blue empty circles to the \textbf{bitwise} one. Lower panel: number of branched warps divided by $k$: for the sliced and the standard implementations the subsequence $L_1 = 2(2m + 1)$ has a divergent warp for each system.}
\label{fig:branch}
\end{figure}

\begin{figure}[h!]
\includegraphics[scale=1.0]{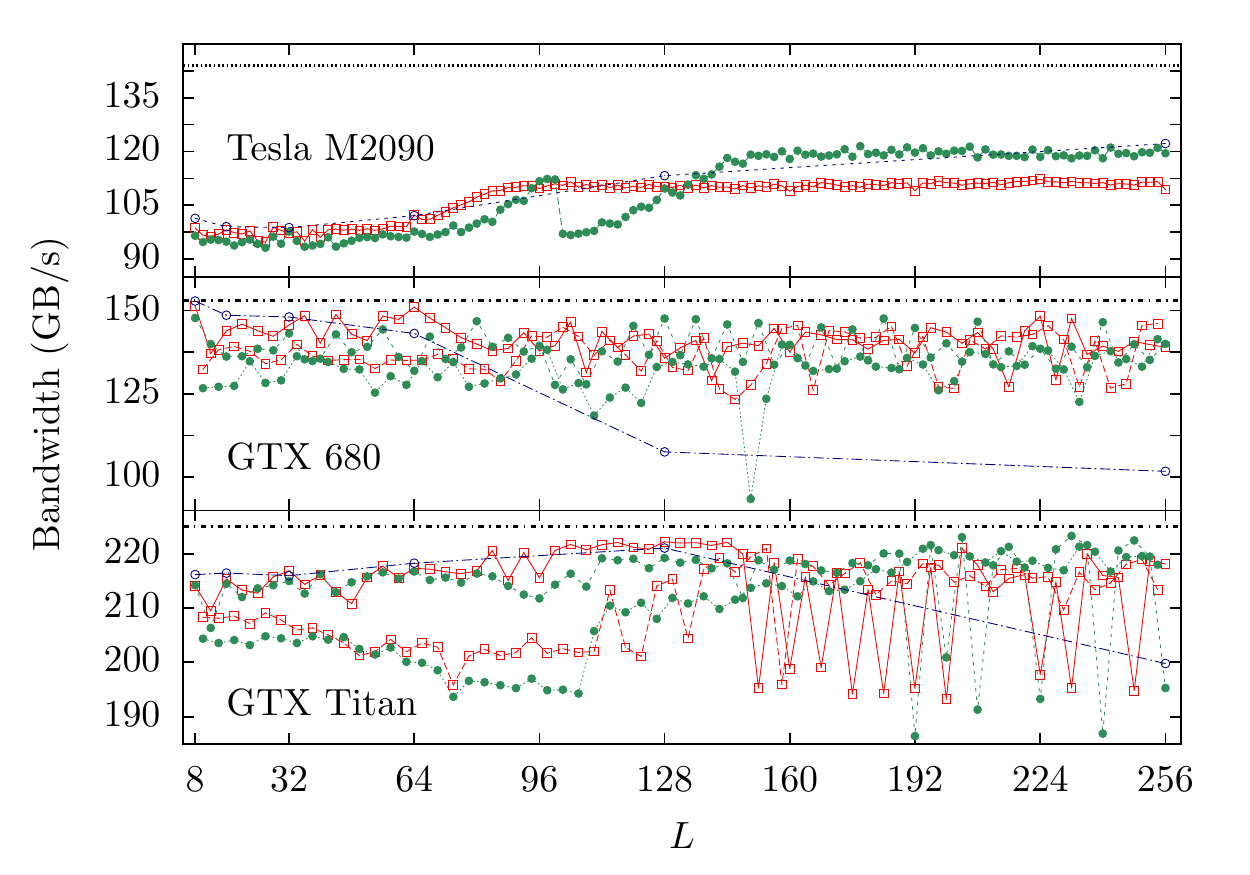}
\caption{Data for the GTX 680 and GTX Titan GPUs. The red empty squares refer to the \textbf{sliced} implementation, the light-green filled circles refer to the \textbf{standard} implementation whereas the blue empty circles to the \textbf{bitwise} one. The horizontal lines are the peak bandwidths as measured with CUDA SDK code.}
\label{fig:bw}
\end{figure}
In Fig. \ref{fig:branch} we report the number of different coded systems $k$ as a function of $L$ and the number of branching warps normalized to $k$: such a ratio gets only two values marking a distinction between the two lattice size subsequences. More details can be found in Appendix B. To complete the analysis for the best performances, we report in Fig. \ref{fig:bw} the results for the bandwidth measures for the best launch configurations. Except for the Tesla M2090, the sliced and the standard algorithms saturate the available bandwidth in the entire range with some fluctuations.

\begin{figure}[h!]
\centering
\subfigure[]{
\includegraphics[scale=0.9]{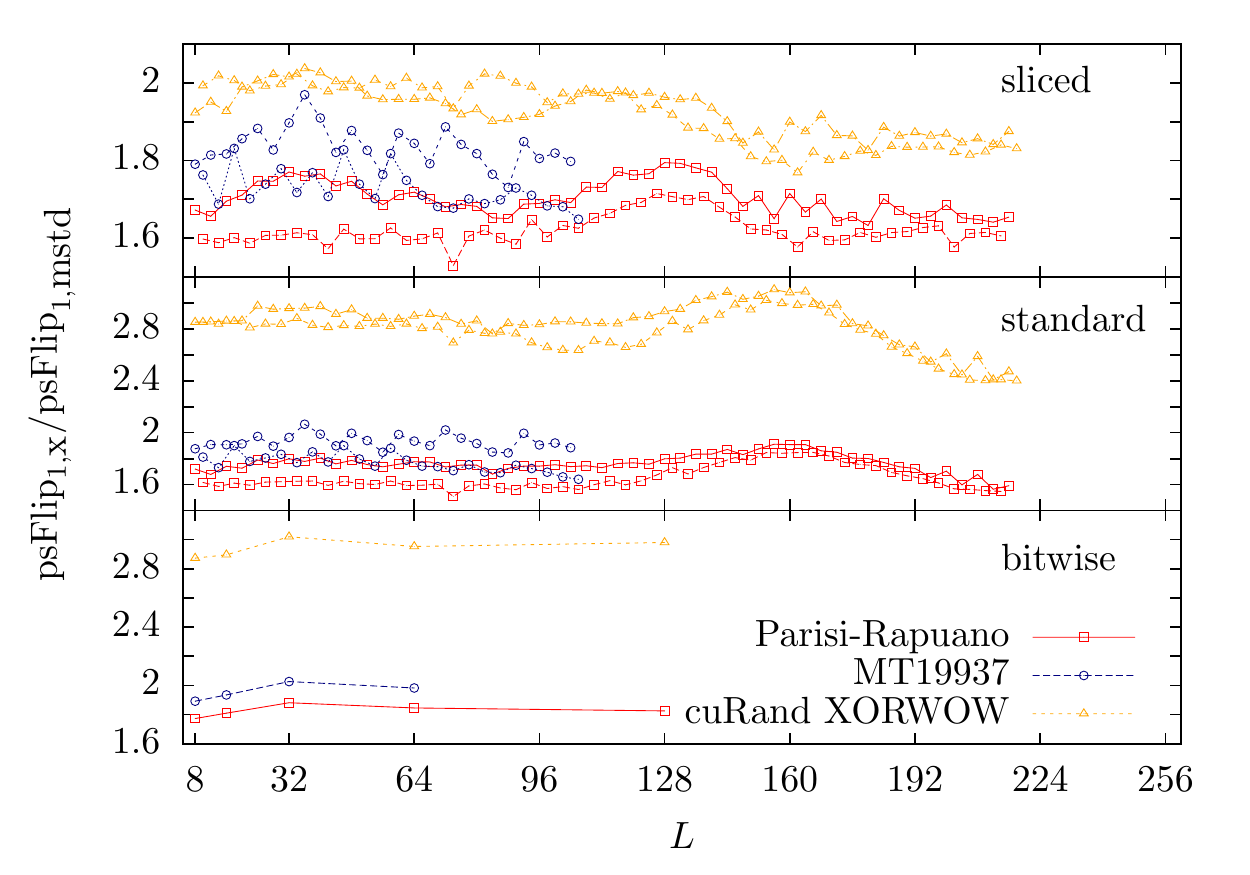}
\label{fig:PRNGRel}
}
\subfigure[]{
\includegraphics[scale=0.9]{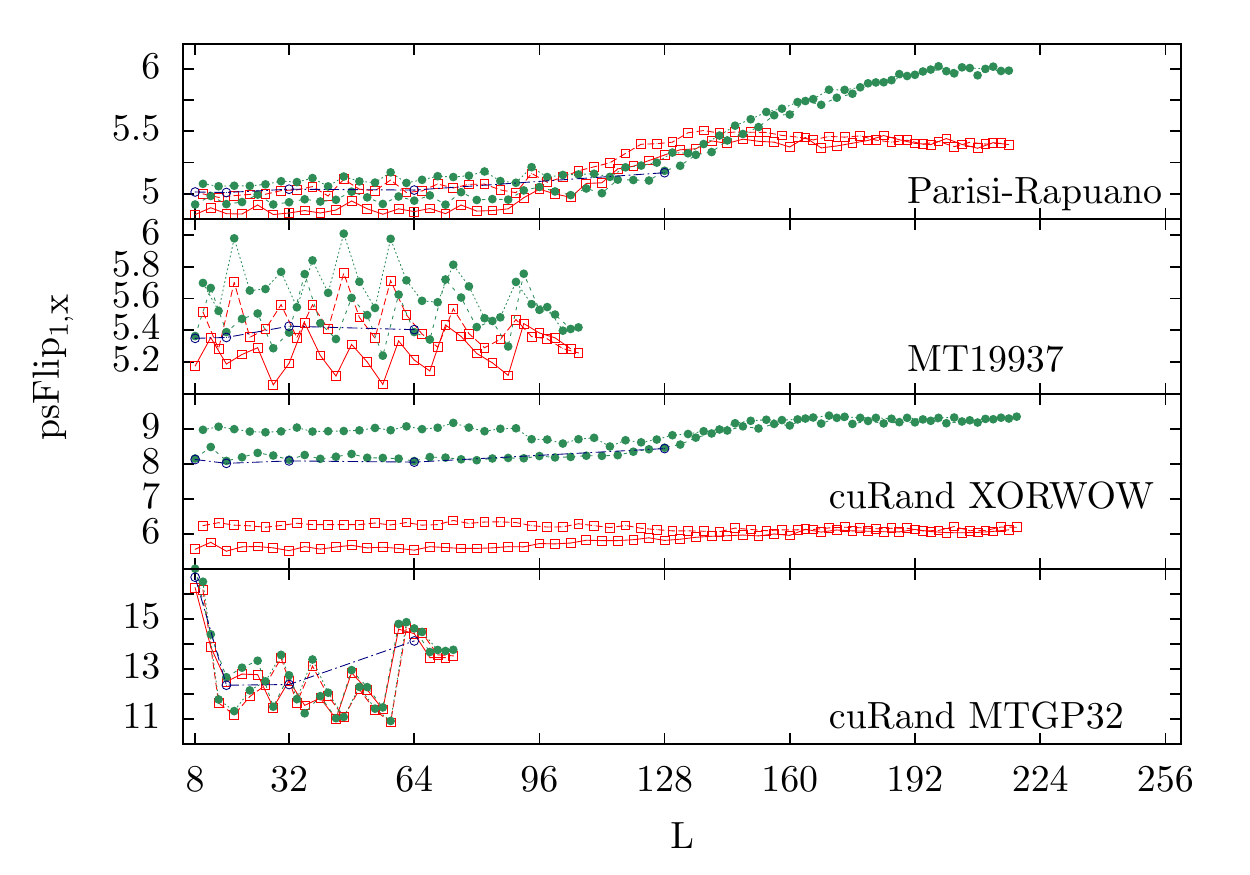}
\label{fig:PRNGAbs}
}
\caption{In figure (a) the $\mbox{psFlip}_{1,\mbox{x}}$ performances normalized to the MINSTD are shown. In figure (b) the abolute performances are reported where the red squares represent the sliced scheme data, the filled green circles and the empty blue circles those of the standard and bitwise schemes respectively. All data refer to the GTX Titan.}
\end{figure}

Let us now examine the results for different PRNGs: our implementations of the Parisi-Rapuano and the usual Mersenne Twister MT19937, together with the cuRand XORWOW (which is the standard cuRand PRNG) and MTGP32 which is a reduced version of the MT19937. While for the first three PRNGs we could perform full benchmarks with the only limitation of the memory usage, for the MTGP32 we could use a maximum number of 200 blocks and a maximum blocks size of 256 threads. As for the number of blocks, this is a limitation of the standard usage which, however, can be by-passed with some effort as reported in the cuRand documentation \cite{cuRandMTGP32DeviceApi}.

Results are reported in Fig. \ref{fig:PRNGRel} and \ref{fig:PRNGAbs}. In Fig. \ref{fig:PRNGRel} we show the values of $\mbox{psFlip}_{1,\mbox{x}}$ normalized to the MINSTD performances $\mbox{psFlip}_{1,\mbox{mstd}}$, which we use as a baseline, for the three different algorithm implementations. In $\mbox{psFlip}_{1,\mbox{x}}$, `x' labels three different PRNGs: Parisi-Rapuano, MT19937 and XORWOW. All data refer to the GTX Titan GPU. It is clear that the lowest ratio for the XORWOW is obtained for the sliced implementation for which it is $\sim 2$ whereas for the standard and bitwise versions the ratio is $\sim 3$. The Parisi-Rapuano and the MT19937 have roughly the same ratio for the three different algorithms.

In Fig. \ref{fig:PRNGAbs} we report the absolute values for $\mbox{psFlip}_{1,\mbox{x}}$. It is possible to see that the performances for our implementations of the Parisi-Rapuano and MT19937 and of cuRand MTGP32 weakly depend on the chosen algorithm, while there is a considerable difference for the XORWOW for which $\mbox{psFlip}_{1,\mbox{xor}}\sim6$ps for the sliced scheme while $\mbox{psFlip}_{1,\mbox{xor}}\sim9$ps for the standard and bitwise implementations. There are two main results emerging from the data:
\begin{itemize}
\item the standard cuRAND XORWOW performs slower than our best-quality PRNG, the MT19937;
\item the sliced scheme is more robust with respect to a change in the memory bandwidth load.
\end{itemize}
The first point can be easily understood by considering that the data structure of the cuRand XORWOW PRNG has a size of 48 bytes: each 128 byte transaction, which is served from the L2 cache, only loads the data needed by two threads, so that we need roughly 16 memory transactions for a warp to be ready, whereas in our approach we only need $O(1)$ memory transactions, \emph{e.g.} 3 for the Parisi-Rapuano and the MT19937. Indeed, also the MTGP32 follows a similar pattern because every thread in a warp loads in the shared memory one entry of the state. The strategy used for the XORWOW implementation is not adequate for intense memory usage algorithms.

As for the second point this should be a proof that the memory alignment given by the sliced scheme is better suited for second hits in the caches: indeed the amount of needed data transfers is the same for the three schemes in the XORWOW case but for the sliced scheme there is a $\sim 33\%$ gain with respect to the standard and bitwise schemes.

As a final remark, the MTGP32 performs from 3 to 5 times worse than the MINSTD implementation. We stress that this result is strongly influenced by some limitations of the cuRand implementation which, however, can be softened with some further work.

\section{Multi-GPU Implementation}
As far as we know, there are just a few works showing strong scaling results for spin systems \cite{Block20101549,Bernaschi20121416, Bernaschi2013250}. We chose to adopt the same technique proposed in \cite{Bernaschi20121416, Bernaschi2013250} where the partitioning is performed along the $z'$-axis of the system. All communications among nodes are handled by MPI and the overlap between calculations and communications is achieved by using CUDA streams. We keep the single-GPU version flexibility for a customary number of spins per thread and coded systems $k$. A priori, it should not be taken for granted that the bulk update, executed on one CUDA stream, can mask the boundary update and data copy/transfer, executed on the other stream, since the algebraic intensity of the algorithm is rather low.
\begin{figure}[h!]
\includegraphics[scale=1.0]{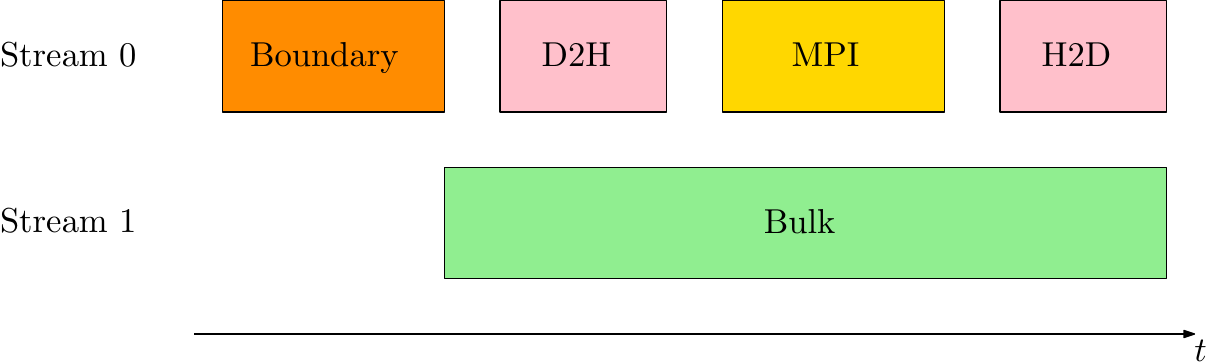}
\caption{Scheme representing the multi-GPU strategy leveraging CUDA streams. Here `Boundary' and `Bulk' represent two kernels launched on the same GPU. After the boundary update on the stream 0 an asynchronous `D2H' device-to-host copy of the only one-coloured boundary is performed, then `MPI' handles the one-directional boundary exchange between nodes and an asynchronous `H2D' host-to-device memory copy updates the boundary spins.}
\label{fig:MGPUSliced}
\end{figure}

The multi-GPU version of the sliced Kernel is rather different from the one using the standard checkerboard scheme \cite{Bernaschi20121416, Bernaschi2013250} since the disposition of colours in the cubic lattice is different. At fixed $z'$ value spins are one-coloured so that for every partition of the system the lowermost plane is always red whereas the highermost one is always blue. This means that when updating red spins the only boundary coincides with the lowermost red plane or with the highermost blue one when updating the blue spins. Hence, the communication between the nodes goes in the downward direction for red spins and in the upward one for the blue spins: there is no need for all nodes to communicate with all nearest neighbours after a colour update. To-be-sent boundary spins are stored in the bulk array and copied to an auxiliary buffer by the same kernel that performs the update. To-be-recieved boundary spins are stored in a separate array, bound to a texture, which is just read when updating the spins of the other colour. This scheme automatically handles the $z'$-axis periodic boundary conditions and reduces the number of intra-node communications. In Fig. \ref{fig:MGPUSliced} we report a depiction of the multi-GPU sliced scheme.

This is an interesting property which might be of use in cases where the amount of data to transfer is low and the latency time is comparable to the data-exchange time. Then, one would expect to have a significant speed-up in the communication.

\begin{figure}[h!]
\includegraphics[scale=0.85]{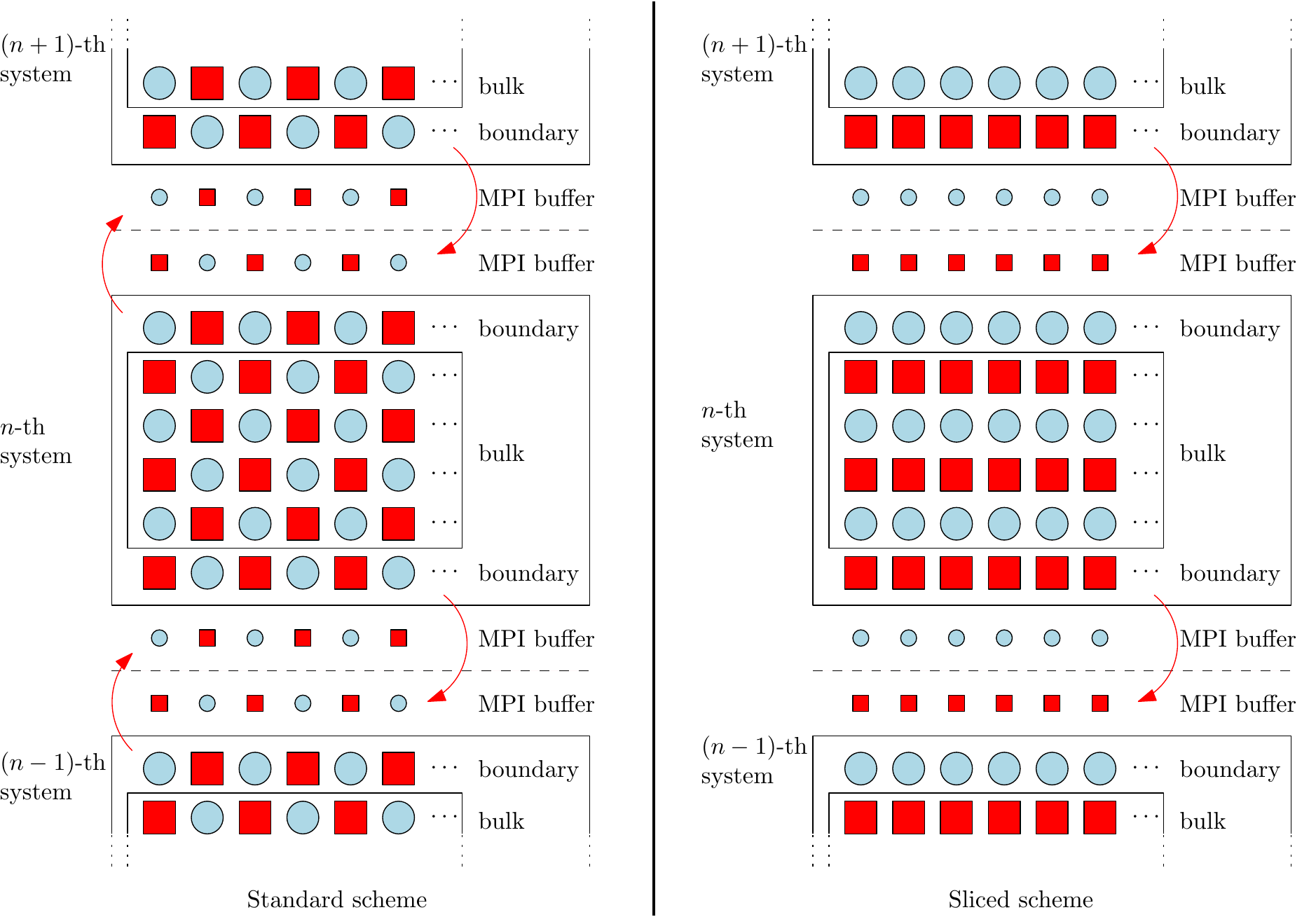}
\caption{A depiction of the standard (on the left) and the sliced (on the right) checkerboard schemes for the multi-GPU version. For the update of red spins one needs to update the bulk and the boundaries of each system partition. The standard scheme has two-coloured boundaries while for the sliced scheme these are one-colured: communication (red arrows) must be two-ways for the standard implementation whereas it is only one-directional for the sliced scheme. Clearly, in both cases the same amount of data is transferred.}
\label{fig:MGPUSliced}
\end{figure}

\subsection{Results}
\begin{figure}[h!]
\centering
\subfigure[]{
\includegraphics[scale=1.0]{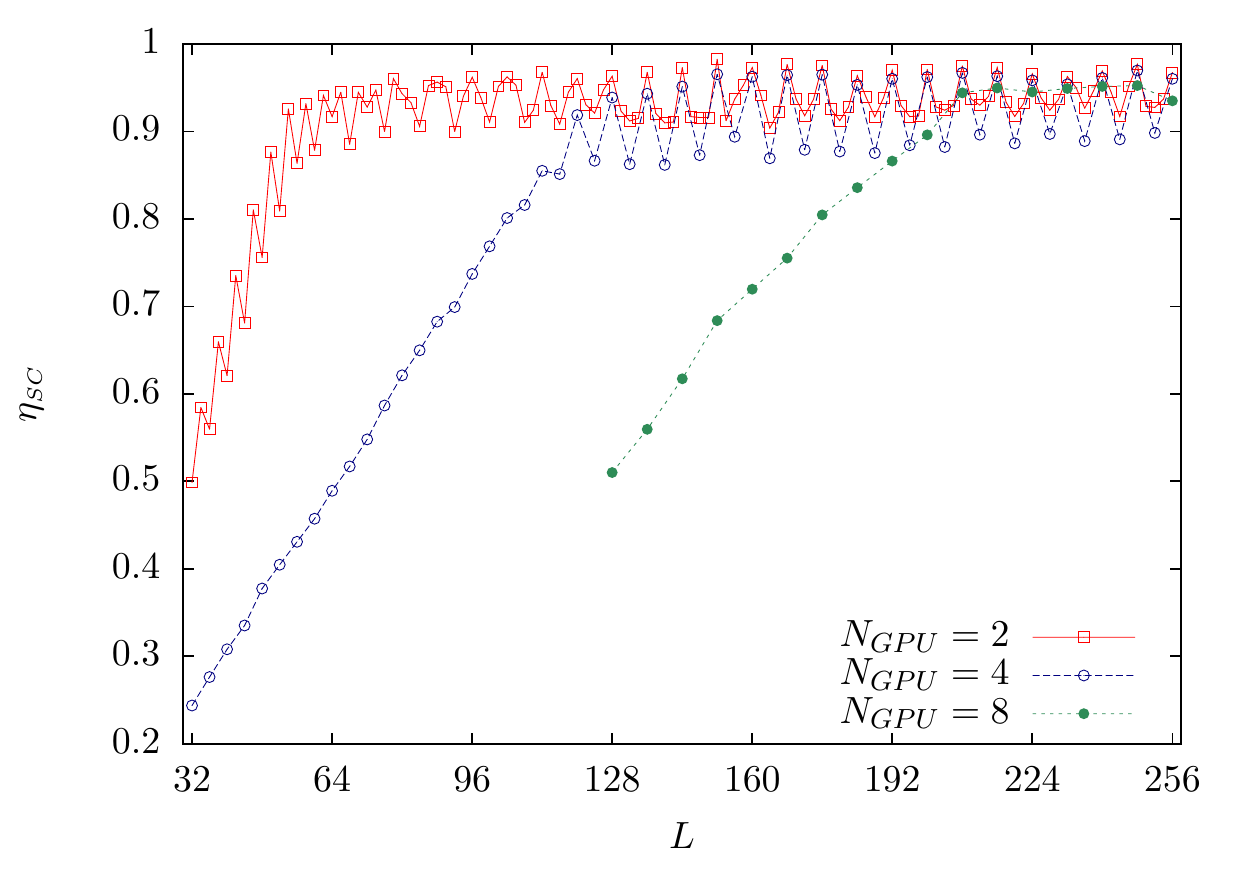}
\label{fig:MGPUeff}
}
\subfigure[]{
\includegraphics[scale=1.0]{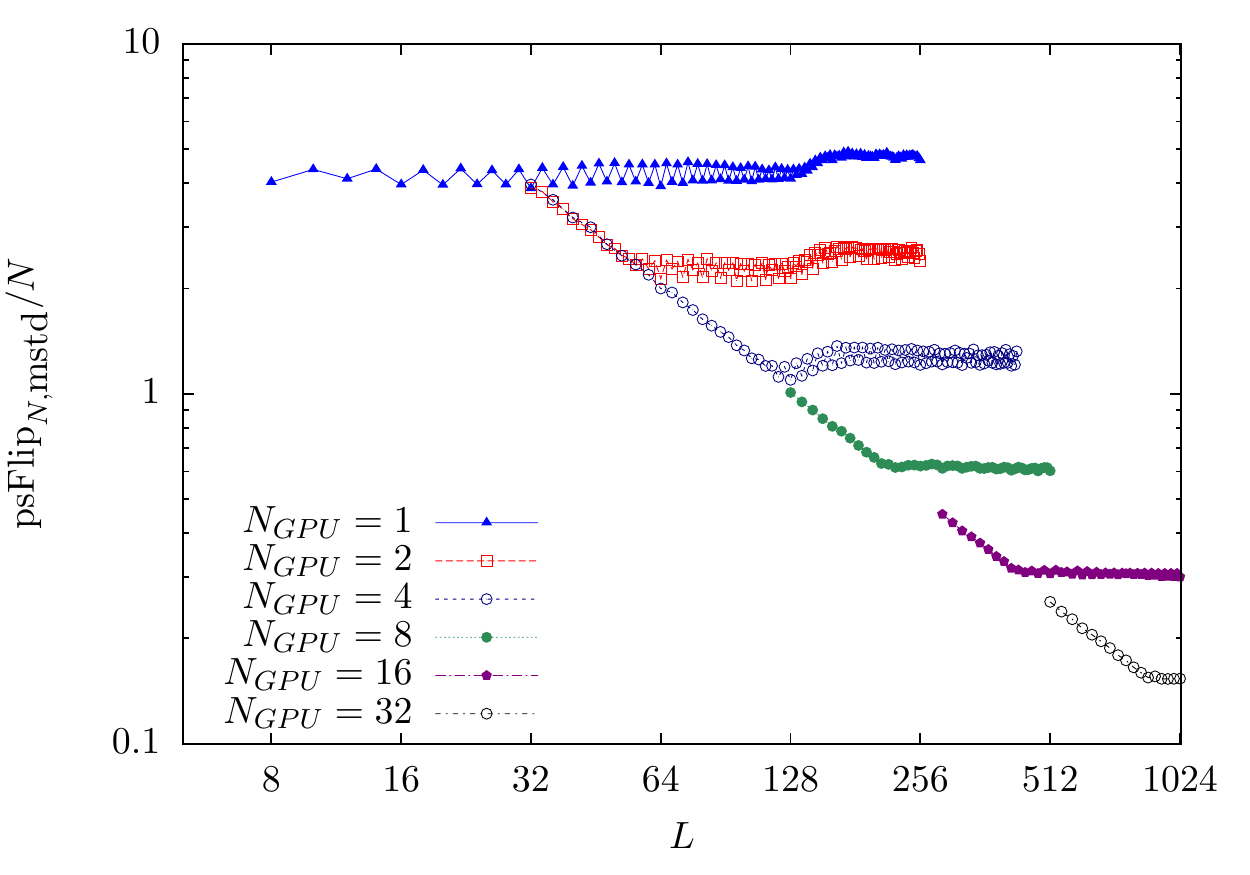}
\label{fig:MGPUpsFlip}
}
\caption{In Fig. (a) the strong-scaling efficiency $\eta_{\scriptscriptstyle{SC}}$ is reported for different numbers of GPUs. In Fig. (b) the multi-GPU system performances are shown. A power-law behaviour as $\mbox{psFlip}_{\mbox{multi},\mbox{mstd}} \sim L^{-1}$ is noticeable.}
\end{figure}

Let us now discuss the results we obtained for the multi-GPU implementation of the three-dimensional Edwards-Anderson model. Given the definition \eqref{eq:psFlip} of $\mbox{psFlip}_{N,\mbox{x}}$, it clearly appears that for $N>1$ we consider the time spent by a single GPU on its own system partition rather than the wall-clock time spent by the $N$ GPUs as a whole. However, the strong-scaling efficiency $\eta_{\scriptscriptstyle{SC}}$ is directly defined as
\begin{equation}
\eta_{\scriptscriptstyle{SC}} = \frac{\mbox{psFlip}_{1,\mbox{x}}}{\mbox{psFlip}_{N,\mbox{x}}},
\end{equation}
and thus the performance referred to the multi-GPU system as a whole is defined as
\begin{equation}
  \mbox{psFlip}_{\mbox{multi},\mbox{x}} = \mbox{psFlip}_{N,\mbox{x}}/N,
\end{equation}
allowing to recover the usual strong-scaling efficiency definition.
\begin{equation}
\eta_{\scriptscriptstyle{SC}} = \frac{\mbox{psFlip}_{1,\mbox{x}}}{\mbox{psFlip}_{N,\mbox{x}}} = \frac{\mbox{psFlip}_{1,\mbox{x}}}{N\cdot \mbox{psFlip}_{\mbox{multi},\mbox{x}}}.
\end{equation}
All data have been gathered on the Piz Daint Supercomputer which uses Tesla K20x GPUs \cite{daint}. In Fig. \ref{fig:MGPUeff} we report the strong scaling efficiency up to 8 GPUs. Indeed, the saturation efficiency is remarkable, $\eta_{\scriptscriptstyle{SC}} \gtrsim 0.9$, although the more the GPUs the further in terms of lattice size $L$ one needs to go to reach a stable regime. Nonetheless up to 8 GPUs the algorithm practically scales linearly with the number of GPUs.

In Fig. \ref{fig:MGPUpsFlip} where we show the values of $\mbox{psFlip}_{\mbox{multi},\mbox{mstd}}$, hence considering $N$ GPUs as a single system, the linear scaling in $N$ is clearly visible for any number of GPUs. We obtain very good results in absolute terms: for $N=2$ in the range from $L=64$ to $L=128$ we have almost stable performances at $2\,\mbox{ps} <\mbox{psFlip}_{\mbox{multi},\mbox{mstd}}<3\,\mbox{ps}$ which is of interest for real scientific applications.

\begin{figure}[h!]
\includegraphics[scale=1.0]{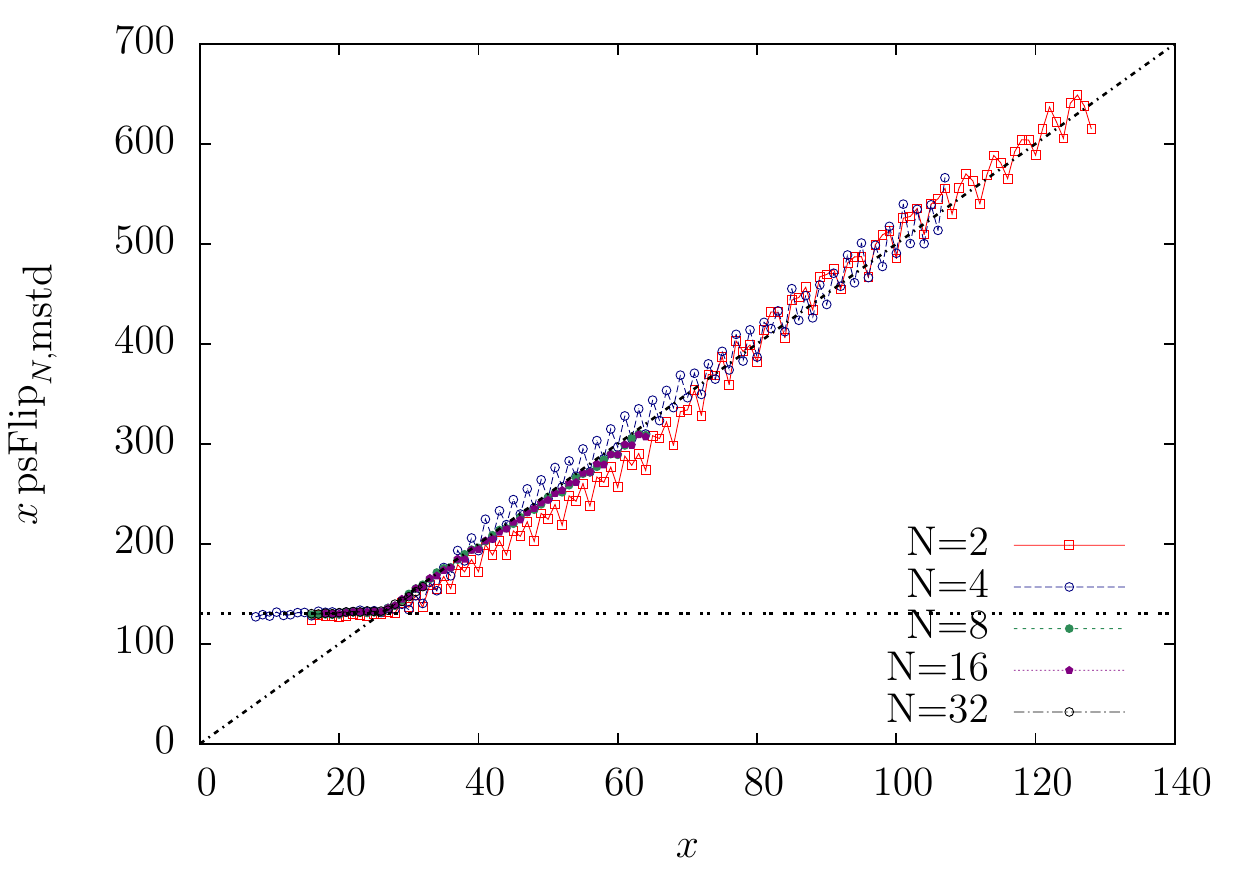}
\caption{Scaling plot for the performances of the multi-GPU system. The initial constant value indicates a scaling for the sweep wall-clock time as $L^2$, whereas for the linear growth in $x$ the sweep time scales as $L^3$ signaling a cross-over from a communication dominated to a bulk calculations dominated regime. The data collapse is possible for the high quality of the inter-node communication.}
\label{fig:MGPUslopes}
\end{figure}
Lastly, we want to pay some attention to the power-law behaviour visible in \ref{fig:MGPUpsFlip}. It is easy to determine that, roughly, performances scale as
\begin{equation}
  \mbox{psFlip}_{\mbox{multi},\mbox{mstd}} \sim L^{-1}.
\end{equation}
Now, looking at the definition \eqref{eq:psFlip}, it is easy to derive that
\begin{equation}
  \frac{L}{N}\,\mbox{psFlip}_{N,\mbox{x}}(L, k) \propto t_{\mbox{sw}} \left( \frac{L}{N}\right)^{-2},
\end{equation}
hence, defining the rescaled variable $x=L/N$ we can plot $x\,\mbox{psFlip}_{N,\mbox{x}}$ as a function of $x$. The result is shown in Fig. \ref{fig:MGPUslopes}. Indeed, we can see that data for every considered value of $N$ collapse on the same curve, proving that $x$ is a good scaling variable. From the plot two distinct regimes are visible: a first one where data lie on a horizontal line and a second one where they grow linearly in $x$. In the first regime the sweep wall clock time grows as $t_{\mbox{sw}} \sim L^2$, \emph{i.e.} the boundary communication, which scales as the system area, dominates. In the second regime $t_{\mbox{sw}} \sim L^3$, which means that the wall-clock time is dominated by the bulk update task which is then able to mask the communication between the nodes.

Indeed, such a good data scaling and collapse shows the stability of the communication offered by the Piz Daint Supercomputer based on Aries routing, communications ASIC, and Dragonfly network topology \cite{kim2008technology,alverson2012cray}. It is possible to use this kind of analysis in order to measure the communication infrastructure quality.
\section{Conclusions}
In this work we have studied different strategies for the implementation of the Metropolis dissipative dynamics for the three-dimensional Edwards-Anderson bimodal spin glass. We proposed new access patterns for both the cubic stencil data structure and lagged-Fibonacci-like PRNGs. We showed, comparing different GPUs and different algorithm implementations that it is possible to obtain stable performances on a wide range of lattice sizes, $8\leq L \leq 256$. For some GPUs our new \textbf{sliced} scheme performs slightly better than the other schemes, for the version which uses the MINSTD PRNG. However, the sliced scheme performs always better for large sizes $L$ and when the data transfer load is increased using more complex PRNGs. In particular the sliced scheme gains roughly the 30\% over standard implementation for the cuRand XORWOW. As for the comparison of different PRNGs we showed that our implementation of the full Mersenne-Twister MT19937 performs better than the standard cuRand XORWOW thus indicating a new implementation strategy for PRNGs which turns out to be very efficient for memory bandwidth demanding algorithms. Indeed, the MT19937 performs only 70\% worse than the MINSTD congruential PRNG with our approach.

Of course at the basis of such results there is the possibility of using the asynchronous multispin-coding (AMSC) technique which allows us to store one spin of 32 different systems in a word. We explained how this technique is implemented in our case.

In terms of single GPU we showed that it is possible to obtain performances comparable to those of dedicated FPGA hardware \cite{Janus2012Tech} although one should be careful in this respect. However, single GPU performances are enough to obtain competitive results for critical parameters estimations using the out-of-equilibrium relaxation regime \cite{LPP-14}.

Furthermore, we explored the multi-GPU version of the sliced scheme which presents the intriguing feature of halving the number of MPI data transactions while, obviously, keeping the total amount of data transfer fixed. We showed that a very high strong-scaling efficiency can be reached leading to scientifically interesting performances in the range $64\leq L \leq 128$.

Many of these result can be extended and reused outside the statistical mechanics domain since they involve cubic lattice discretization, along with their multi-GPU extension, and high quality random numbers PRNGs implementations.

\section{Acknowledgements}
We gratefully acknowledge the support of 
NVIDIA Corporation with the donation of the GTX Titan GPUs used for 
this research.
The research leading to these results has received funding from the European Research Council under the European Union's Seventh Framework Programme (FP7/2007-2013) / ERC grant agreement number [247328]. We also wish to acknowledge Dr. M. Spera and Prof. A. Pelissetto for useful discussion.

\appendix
\section{Grid parameters and periodic boundary conditions}
We now list some details for grid configurations of non-Grid and Grid kernels.
The kernel launch parameters for the first case are defined as follows
\begin{verbatim}
   dim3 block(blockSize,1,1);
   int fitGrid = (V/2/s + blockSize - 1)/blockSize;
   dim3 grid(fitGrid, k, 1);
\end{verbatim}
where $\texttt{s} = s$ is the number of spins per thread and $\texttt{blockSize} = 32n$, \emph{i.e.} a multiple of the warp size. For Grid kernels the launch parameters are:
\begin{verbatim}
   dim3 blockG(L, l, 1);
   dim3 gridG(A/(blockG.x*blockG.y), L/2, k);
\end{verbatim}
where \texttt{A=L*L} and \texttt{l} is the number of lines of a single plane updated by a thread block. We highlight that for the latter case we use \texttt{threadIdx.x}, \texttt{threadIdx.y} and \texttt{blockIdx.y} as $x$, $y$ and $z$ indices respectively.

Let us now discuss some implementations details for periodic boundary conditions. The nearest neighbours indices are always calculated from the one-dimensional index of the spin that is updated. Here we report the calculation which are the same for  the sliced and sliced-Grid kernels in order to be as clear as possible
\begin{verbatim}
  int smz = i + (SM(z - 1, d_hL) - z)*d_A;
  int spy = smz + (SP(y + 1, d_L) - y)*d_L;
  int smy = i + (SM(y - 1, d_L) - y)*d_L;
  int smx = spy - x + SM(x - 1, d_L);
  int spx = smy - x + SP(x + 1, d_L);
\end{verbatim}
where \texttt{i = kk + off}, with $\texttt{kk} < V/2$ and \texttt{off = blockId.y,z*d\_hV} being a disorder offset (with $\texttt{d\_hV} = V/2$). We have implicitly set \texttt{spz = i}. In order to avoid the modulus operation enforcing the periodic boundary conditions we defined the macros \texttt{SM} and \texttt{SP} which read
\begin{verbatim}
#define SP(a, m) (a&(~(-(a >= m))))
#define SM(a, m) (a+((-(a < 0))&m))
\end{verbatim}
Let us briefly comment the definition of \texttt{SP}: if \texttt{a >= m} it evaluates to \texttt{1} then
\begin{verbatim}
(~(-(a >= m))) = 0x00000000
\end{verbatim}
\emph{i.e.} all bits set to zero, otherwise one has
\begin{verbatim}
(~(-(a >= m))) = 0xffffffff
\end{verbatim}
\emph{i.e.} all bits set to one. The macro \texttt{SM} is completely analogous. Hence we have reproduced the periodic boundary conditions since \texttt{SP(m + 1, m) = 0} and \texttt{SM(-1, m) = m - 1}.

We remark that we had to define another macro \texttt{SMM} for the Grid version in order to handle the fact that \texttt{threadIdx} and \texttt{blockIdx} variables are unsigned integers.

\section{Even $L$ subsequences and warp branchings}
In Fig.\ref{fig:branch} we report the optimal values of the number of coded systems $k$ as a function of the lattice size which we can see decreases roughly in a power-law fashion. In particular we also show the ratio between the number of branching warps and $k$: for lattices belonging to the subsequence $L_1 = 2(2m + 1)$ this ratio is always equal to one, whereas for the subsequence $L_0 = 4m$ it is always equal to zero. This result is explained by the fact that half of the volume of a lattice $V_i/2 = L_i^3/2$, \emph{i.e.} all the one-coloured spin, is always a multiple of the warp size for the even subsequence $V_0/2 \propto 32$ whereas it is not so for the odd subsequence $V_1/2$,
\begin{equation}
  \frac{V_0}{2} = \frac{(4m)^3}{2} = 32m^3,\qquad\frac{V_1}{2} = \frac{[2(2m + 1)]^3}{2} = 4(2m + 1)^3.
\end{equation}
In order to prove this let us look if there exist a value of $m$ for which $V_1/2$ is a multiple of the warp size
\begin{equation}
  \frac{V_1}{2} = 4(2m + 1)^3 = 32n,\qquad 2m + 1 = 2 n^{1/3},\qquad m = n^{1/3} + \frac{1}{2},
\end{equation}
which has integer solutions for $m$ for non-integer $n$. Since we are looking for integer values of $n$, this proves the previous assertion. Hence, the subsequence of cubic lattices of linear size $L_1$ is intrinsically uncommensurate to the actual warp size which is characteristic of the CUDA framework. Thus, as long as the warp size is fixed to the actual value, there will always be warp branchings for checkerboard algorithms updating one colour at the time. Indeed, this result is correlated to the fact that the two subsequences $L_0$ and $L_1$ have different performances, but does not provide a full explanation.

\bibliographystyle{ieeetr}
\bibliography{biblio}
\end{document}